\documentclass[11pt,a4paper]{article}
\pdfoutput=1

\usepackage{jheppub}
\usepackage[T1]{fontenc}
\usepackage{extarrows}

\usepackage{todonotes}

\usepackage{xcolor}

\graphicspath{{Figures/}}

\title{U(1) Current Diffusion in Gauss-Bonnet Holography}

\author[a]{A. Naveena Kumara,}
\author[b]{Shreyas Punacha,}
\author[c]{and Kartheek Hegde}
\affiliation[a]{Rudjer Bo\v{s}kovi\'c Institute, Bijeni\v cka  c.54, HR-10002 Zagreb, Croatia}
\affiliation[b]{Department of Oral Health Sciences, School of Dentistry, University of Washington, Seattle, WA 98195, USA}
\affiliation[c]{Department of Physics, National Institute of Technology Karnataka, Surathkal 575 025, India}

\emailAdd{naviphysics@gmail.com}
\emailAdd{nathith@irb.hr}
\emailAdd{shreyas4@uw.edu}
\emailAdd{hegde.kartheek@gmail.com}

\abstract{We study holographic \(U(1)\) charge transport using a probe Maxwell field in a five-dimensional Gauss–Bonnet–AdS black-brane background. Within the all-order linearised fluid/gravity framework, we derive the off-shell constitutive relation and determine the diffusion function and electric and magnetic conductivities analytically in the hydrodynamic regime and numerically at generic frequency and momentum. To first order in the Gauss–Bonnet coupling, the leading diffusion constant remains unchanged in our normalization, while its frequency dependence and the electric conductivity receive nontrivial corrections. The deformation also generates a nonvanishing leading magnetic conductivity. Beyond the hydrodynamic regime, it modifies the transport profiles while preserving the damped behaviour of the diffusion function. The corresponding memory kernel vanishes at negative times, supporting a retarded constitutive response.}


\begin{document}
\maketitle

\flushbottom

\section{Introduction}

Hydrodynamics is the effective language of many-body physics at wavelengths long compared to microscopic relaxation scales. Its variables are not the underlying quantum fields themselves, but the expectation values of conserved currents and the slowly varying thermodynamic data that organise near-equilibrium evolution. In this sense, hydrodynamics is less a model than a universal coarse-grained framework: once the relevant symmetries and conservation laws are specified, the remaining microscopic input is encoded in constitutive relations and the transport data that accompany them. The practical difficulty, of course, is that these transport coefficients are often most interesting precisely in regimes where microscopic calculations are hardest. Kubo formulae provide the formal bridge between transport and real-time correlation functions, but they do not by themselves solve the strong-coupling problem~\cite{Kubo:1957mj, forster2018hydrodynamic, Kovtun:2012rj, Romatschke:2017ejr}.

Gauge/gravity duality offers a remarkably efficient route around this obstacle. In its standard form, the AdS/CFT correspondence relates a strongly coupled large-$N$ quantum field theory to a weakly coupled gravitational theory in asymptotically anti-de Sitter spacetime~\cite{Maldacena:1997re, Gubser:1998bc, Witten:1998qj}. Real-time transport is then translated into a problem of classical wave propagation in black-brane geometries, with boundary observables extracted by holographic renormalization~\cite{Bianchi:2001kw, Son:2007vk}. What makes this framework especially compelling for hydrodynamics is that the long-time, long-distance regime of the boundary theory is encoded in near-horizon bulk dynamics. The earliest holographic computations of shear viscosity and charge diffusion already made clear that dissipative transport in strongly coupled plasmas can be read directly from black-hole physics~\cite{Policastro:2001yc, Policastro:2002se, Kovtun:2003wp, Kovtun:2004de, Iqbal:2008by}. This observation was elevated to a systematic correspondence once it became understood that slowly varying black-brane solutions furnish the dual description of relativistic fluid flow~\cite{Bhattacharyya:2007vjd, Rangamani:2009xk, Janik:2005zt, Banerjee:2008th, Erdmenger:2008rm, Haack:2008xx}.

The constitutive side of relativistic hydrodynamics is, however, more subtle than the familiar non-relativistic intuition might suggest. Truncating the derivative expansion at first order leads to parabolic evolution equations, and therefore to acausal signal propagation. Diffusion theory already exhibits this problem in its simplest form, and relativistic Navier--Stokes inherits it in a broader setting. Müller--Israel--Stewart theory and related second-order formulations were developed to repair this deficiency by introducing finite relaxation scales and promoting dissipative fluxes to transient degrees of freedom/ additional transient dynamical variables~\cite{Israel:1976tn, Stewart1977, Israel:1979wp, Baier:2007ix}. These frameworks have been indispensable in phenomenological applications, but they also make clear that causality is not a property of any arbitrary truncation. Rather, it is tied to the analytic structure of the fully resummed constitutive relation. Recent progress on stable and causal formulations of relativistic hydrodynamics has sharpened this lesson further and reinforced the view that relativistic dissipation is best understood beyond the naive finite-order gradient expansion~\cite{Kovtun:2019hdm, Romatschke:2017ejr}.

From this perspective, all-order hydrodynamics is not merely an aesthetic improvement over low-order truncations; it is the natural language for causal constitutive relations. The resummation of higher-order gradient corrections was investigated by Lublinsky and Shuryak~\cite{Lublinsky:2007mm, Lublinsky:2009kv,Lublinsky:2011cw}, who developed an all-order formulation of linearised relativistic hydrodynamics in terms of momentum-dependent transport functions and explored its implications for dissipative dynamics and causality. In Fourier space, the usual transport coefficients are replaced by momentum-dependent transport functions. In real time, the same information is encoded in memory kernels, so that the current at time $t$ depends on the past history of the system rather than only on instantaneous gradients. The general philosophy is older than holography and has deep roots in nonequilibrium statistical mechanics~\cite{Kubo:1957mj, Mori:1965oqj, forster2018hydrodynamic}. What holography adds is a first-principles setting in which these transport functions can be computed nonperturbatively for strongly coupled theories. In the stress-tensor sector, the linearized fluid/gravity program made this point especially concrete by resumming all derivative corrections into viscosity functions, whose inverse Fourier transforms define causal memory kernels~\cite{Bu:2014sia, Bu:2014ena, Bu:2015ika, Bu:2015bwa}. More broadly, recent studies of divergent gradient series, Borel structure, and hydrodynamic attractors have emphasized that the derivative expansion should be viewed as an asymptotic organizing principle whose physical content is completed in many holographic and kinetic-theory examples after suitable resummation~\cite{Heller:2015dha, Romatschke:2017vte}.

The conserved-current sector provides an especially clean arena in which to explore these issues. Once a global $U(1)$ symmetry is present, charge conservation must be supplemented by a constitutive relation for the associated current. Even in the simplest neutral medium, the familiar diffusion law is already acausal if interpreted as an exact relativistic statement. The natural completion is to allow the diffusion constant to become a function of frequency and momentum, and to include the response to external electromagnetic sources through generalized conductivities. In a holographic setting, this problem is encoded in the dynamics of a bulk Maxwell field. Early work on holographic charge transport established the basic relation between black-brane perturbations and diffusion or conductivity~\cite{Policastro:2002se, Kovtun:2003wp, Son:2007vk}. Probe-brane constructions then broadened the class of charged sectors that can be studied while keeping the analysis technically controlled~\cite{Karch:2002sh, Mas:2008qs}. A particularly useful development was the off-shell formulation of the holographic current, in which the constitutive relation is extracted from the dynamical bulk equations, while current conservation arises separately as a constraint. The off-shell construction of the holographic $U(1)$ current was developed by Bu, Lublinsky and Sharon~\cite{Bu:2015ame}, who formulated the constitutive relation in terms of three all-order transport functions: the diffusion function, electric conductivity and magnetic conductivity.

That formulation has since been extended in several directions. The same methodology has been applied to anomaly-induced transport, where the current sector becomes substantially richer due to chiral magnetic, chiral separation, and related effects~\cite{Bu:2016oba, Bu:2016vum, Bu:2018psl, Bu:2018drd, Bu:2019mow, Baggioli:2024zfq, Lublinsky:2026wvt}. More recently, Schwinger--Keldysh holography has provided an all-order effective action for charge diffusion, making it possible to incorporate fluctuations and colored noise within a holographic description that unifies dissipation and stochasticity~\cite{Bu:2020jfo, Crossley:2015evo, Liu:2018kfw, Haehl:2014zda}. These developments make clear that the current sector is not a minor extension of fluid dynamics but a flexible testing ground for questions of causality, nonlocal response, effective actions, and the interplay between hydrodynamic and nonhydrodynamic modes.

A second line of motivation comes from higher-derivative gravity. Einstein gravity captures the universal large-$N$, large-coupling limit of a wide class of holographic theories, but finite-coupling and $1/N$ effects generically induce higher-curvature corrections in the bulk. Among the simplest and most widely studied deformations is Gauss--Bonnet gravity, which preserves second-order metric equations while modifying boundary transport and the relation between the two central charges of the dual conformal field theory~\cite{Brigante:2007nu, Brigante:2008gz, Buchel:2009tt, Kats:2007mq}. In the stress-tensor sector, Gauss--Bonnet gravity became a standard laboratory for understanding how higher-derivative corrections shift transport coefficients, deform dispersion relations, and constrain the allowed parameter space through consistency requirements such as microcausality~\cite{Brigante:2008gz, Buchel:2009tt}. This makes Gauss--Bonnet gravity particularly well suited for asking how robust the constitutive structures found in Einstein gravity remain once the bulk dynamics is deformed away from the universal limit.

The current sector in a Gauss--Bonnet background is therefore a natural next step. It allows one to isolate, in a controlled probe setting, how higher-curvature corrections modify charge transport beyond the lowest hydrodynamic order. Existing studies of Maxwell dynamics in Gauss--Bonnet backgrounds have already shown that even at the level of conventional hydrodynamics the current and conductivity sector is sensitive to the higher-derivative deformation~\cite{Hu:2011ze, Hu:2011qa}. What remains especially interesting is the all-order problem: how the diffusion function and generalized conductivities are reorganized once one keeps the full derivative dependence rather than truncating at the first few gradients. This question is conceptually sharper than merely asking how a few transport coefficients shift. It probes whether the causal, nonlocal constitutive structure familiar from Einstein holography persists in the presence of controlled higher-curvature corrections, and how such corrections reshape the analytic response of the boundary theory.

Recent developments in relativistic hydrodynamics make this question timely from another angle as well. The modern viewpoint increasingly emphasizes hydrodynamics as an effective theory constrained by symmetry, analyticity, and nonequilibrium consistency rather than as a simple low-order gradient expansion. This shift is visible in several directions: the resurgence-based study of hydrodynamic asymptotics and attractors~\cite{Heller:2015dha, Romatschke:2017vte}, the construction of Schwinger--Keldysh effective actions for dissipative systems~\cite{Crossley:2015evo, Liu:2018kfw, Haehl:2014zda}, the renewed analysis of stable and causal relativistic formulations~\cite{Kovtun:2019hdm}, and the extension of hydrodynamic thinking to generalized global symmetries and magnetohydrodynamic-type effective theories~\cite{Grozdanov:2017kyl}. Holographic charge diffusion in a higher-derivative background sits naturally at the intersection of these themes: it is simultaneously a problem in transport theory and the gravitational encoding of causal response.

The aim of the present work is to study the holographic $U(1)$ current in a five-dimensional Gauss--Bonnet black-brane geometry, with the Maxwell field treated in the probe approximation. This setup keeps the charge sector decoupled from energy-momentum backreaction and therefore isolates the constitutive structure of the current itself. Our interest is in the off-shell constitutive relation induced by bulk Maxwell dynamics, its hydrodynamic expansion at small frequency and momentum, and its continuation to all derivative orders in terms of transport functions. Building on the off-shell, all-order $U(1)$ current framework developed in Ref.~\cite{Bu:2015ame}, we extend the analysis to a Gauss-Bonnet black-brane background. The main objective is to determine how higher-curvature corrections modify the diffusion function, electric and magnetic conductivities, and the associated causal memory response.

The paper is organized as follows. In section~\ref{section:two} we introduce the Gauss--Bonnet holographic setup and specify the probe Maxwell dynamics in the black-brane background. In section~\ref{section:three} we derive the off-shell constitutive relation for the boundary current directly from the bulk equations. Section~\ref{section:four} develops the hydrodynamic expansion and identifies the corresponding transport data. We then turn to the analysis of the all-order transport functions and their physical interpretation in section~\ref{section:five}. We conclude with a discussion of the broader implications of higher-derivative corrections for holographic charge transport and for the causal organization of relativistic hydrodynamics.

\section{Gauss-Bonnet Holographic Setup}\label{section:two}

We consider a probe $U(1)$ gauge field propagating in a fixed Gauss-Bonnet–$\text{AdS}_5$ gravitational background. The gauge field is treated in the probe (quenched) approximation: the number of probe branes is assumed to be parametrically smaller than the number of colour branes, so the back-reaction of the Maxwell field on the metric is consistently neglected. This is the setup that arises naturally from probe D-brane constructions such as the D3/D7 system, and it allows us to study charge diffusion and conductivity independently of energy-momentum transport.

The bulk action for the probe Maxwell field is
\begin{equation}\label{maxwell action}
S = \frac{1}{4}\int d^5x\,\sqrt{-g}\,F^{MN}F_{MN} + S_{\textrm{c.t.}},
\end{equation}
where $F_{MN} = \nabla_M A_N - \nabla_N A_M$ is the bulk field strength tensor and $S_{\textrm{c.t.}}$ is the holographic counter-term required for renormalization, to be specified below.\footnote{Throughout, upper-case Latin indices $\{M, N, \ldots\}$ label bulk directions,  lower-case Greek indices $\{\mu,\nu,\ldots\}$ label boundary directions, and   lower-case Latin indices $\{i,j,\ldots\}$ label spatial directions on the boundary.  The determinant of the bulk metric $g_{MN}$ is denoted as $g$.} The bulk Maxwell equations that follow from~(\ref{maxwell action}) are
\begin{equation}\label{Maxwell eq}
\mathrm{EQ}^N := \nabla_M F^{MN} = 0
\;\Longrightarrow\;
\partial_M\!\left(\sqrt{-g}\,F^{MN}\right) = 0.
\end{equation}
We work throughout in radial gauge $A_r = 0$.

The gravitational background is a planar Gauss-Bonnet black brane in asymptotically $\text{AdS}_5$ spacetime. Written in ingoing Eddington-Finkelstein coordinates, the static black brane metric takes the form~\cite{Brihaye:2008kh, Bu:2015bwa}
\begin{equation}\label{gs metric}
ds^2 = 2N_{\#}\,dt\,dr - N_{\#}^2 r^2 f(\mathbf{b}r)\,dt^2 + r^2\,\delta_{ij}\,dx^i dx^j,
\qquad i,j = 1,2,3,
\end{equation}
where $\mathbf{b}$ sets the scale of the black brane. The use of ingoing Eddington-Finkelstein coordinates ensures that the metric is manifestly regular on the event horizon, which is essential for imposing physically sensible in-falling boundary conditions. To linear order in the Gauss-Bonnet coupling $\alpha$, the metric functions are
\begin{equation}\label{metric functions}
N_{\#} = 1 - \alpha + \mathcal{O}(\alpha^2),
\qquad
f(r) = 1 - \frac{1}{r^4} + 2\alpha\!\left(1 + \frac{1}{r^8}\right) + \mathcal{O}(\alpha^2).
\end{equation}
The Gauss-Bonnet coupling $\alpha$ is a dimensionless parameter characterising deviations from pure Einstein gravity; it is related to the difference between the two central charges of the dual CFT and encodes $\mathcal{O}(1/N)$ corrections in the boundary theory. Causality of the dual field theory imposes the constraint~\cite{Brigante:2007nu,Brigante:2008gz,Buchel:2009tt,Buchel:2009sk}
\begin{equation}\label{alpha bound}
-\frac{7}{72} \leq \alpha \leq \frac{9}{200},
\end{equation}
and we shall work within this range throughout. The thermodynamics of Gauss-Bonnet black holes~\cite{Myers:1987yn, Davis:2002gn} yields the event horizon radius $r_\text{h}$ and Hawking temperature $T$,
\begin{equation}\label{horizon}
r_\text{h} = \frac{1 - \alpha}{\mathbf{b}},
\qquad
T = \frac{1 - 2\alpha}{\pi\mathbf{b}}.
\end{equation}
The conformal boundary sits at $r = \infty$. Without loss of generality we henceforth set $\mathbf{b} = 1$, which is equivalent to working in units $\pi T = 1 - 2\alpha$. All dimensionful momenta are understood to be measured in units of $\pi T$.

In an asymptotically AdS spacetime, the on-shell Maxwell action diverges near the conformal boundary and a holographic renormalization procedure is necessary. Following the holographic renormalization procedure for the $U(1)$ gauge field in Eddington-Finkelstein coordinates~\cite{Bu:2015ame}, we determine the
corresponding modifications arising from the Gauss--Bonnet background. An analogous treatment in Fefferman-Graham coordinates can be found in Refs.~\cite{Taylor:2000xw,Bianchi:2001kw,Sahoo:2010sp}.

\paragraph{Near-boundary expansion.} Assuming the Maxwell equations~(\ref{Maxwell eq}) are solved, the bulk gauge field near the conformal boundary $r \to \infty$ admits the asymptotic expansion
\begin{equation}\label{boundary solution}
A_\mu(r, x_\alpha)
= A_\mu^{(0)}(x_\alpha)
+ \frac{A_\mu^{(1)}(x_\alpha)}{r}
+ \frac{A_\mu^{(2)}(x_\alpha)}{r^2}
+ \mathcal{O}\!\left(\frac{\log r^{-2}}{r^3}\right).
\end{equation}
Substituting~(\ref{boundary solution}) back into the Maxwell equations~(\ref{Maxwell eq}) fixes two of the expansion coefficients algebraically,
\begin{equation}\label{subleading terms}
A_\mu^{(1)} = (1-\alpha)\,F^{(0)}_{t\mu},
\qquad
4B_\mu^{(2)} = (1-2\alpha)\,\partial^\nu F^{(0)}_{\mu\nu},
\end{equation}
while $A_\mu^{(0)}$ and $A_\mu^{(2)}$ remain as free integration constants. The leading coefficient $A_\mu^{(0)}$ is identified with the external electromagnetic potential that sources the boundary current, while $A_\mu^{(2)}$ encodes the dynamical response of the medium.

\paragraph{Counter-term and divergence structure.} Inserting the near-boundary expansion~(\ref{boundary solution}) into the regularised Maxwell action~(\ref{maxwell action}) reveals a logarithmic UV divergence near $r = \infty$,
\begin{equation}
S_{\textrm{reg.}}
= \frac{1}{4}\int_{r_\text{h}}^{\Lambda\to\infty}\!dr\int d^4x\,\sqrt{-g}\,F^{MN}F_{MN}
= \frac{1}{4}(1-\alpha)\log\Lambda \int d^4x\,F^{(0)}_{\mu\nu}F^{\mu\nu(0)} + \cdots,
\end{equation}
where the ellipsis denotes finite contributions as $\Lambda\to\infty$. The covariant counter-term that cancels this divergence is
\begin{equation}\label{counter term}
S_{\textrm{c.t.}}
= -\frac{1-\alpha}{4}\log r\int d^4x\,\sqrt{-\gamma}\,F_{\mu\nu}F^{\mu\nu},
\end{equation}
where $\gamma_{\mu\nu}$ is the induced metric on the constant-$r$ hypersurface $\Sigma$,
\begin{equation}
ds^2\big|_\Sigma = \gamma_{\mu\nu}\,dx^\mu dx^\nu 
= - N_{\#}^2 r^2 f(r)\,dt^2 + r^2\,\delta_{ij}\,dx^i dx^j,
\end{equation}
and $\gamma = \det\gamma_{\mu\nu}$. Note that the counter-term inherits the Gauss-Bonnet correction through the overall factor $(1-\alpha)$, which is a direct consequence of the modified near-boundary geometry.

\paragraph{Boundary current.} The boundary current is defined via the variational prescription
\begin{equation}
J^\mu \equiv \lim_{r\to\infty} \frac{\delta S}{\delta A_\mu}.
\end{equation}
Using the bulk Maxwell equations~(\ref{Maxwell eq}), the variation of the full renormalized action evaluates to
\begin{equation}\label{variation}
\delta S
= \int d^4x\,\sqrt{-\gamma}
  \left\{
    \delta A_N\,F^{MN}\,n_M
    + (1-\alpha)\log r\;\delta A_\mu\,\widetilde{\nabla}_\nu F^{\nu\mu}
  \right\},
\end{equation}
where $\widetilde\nabla_\mu$ is the covariant derivative compatible with $\gamma_{\mu\nu}$, and $n_M$ is the outward-pointing unit normal to $\Sigma$,
\begin{equation}
n_M \equiv \frac{\nabla_M r}{\sqrt{g^{AB}\nabla_A r\,\nabla_B r}}.
\end{equation}
The resulting boundary current is
\begin{equation}
J^\mu = \lim_{r\to\infty}
\left\{
  \sqrt{-\gamma}\,F^{M\mu}\,n_M
  + (1-\alpha)\log r\;\sqrt{-\gamma}\,\widetilde\nabla_\nu F^{\nu\mu}
\right\}.
\end{equation}
Re-expressed in terms of the near-boundary expansion coefficients of~(\ref{boundary solution}), this takes the compact form 
\begin{equation}\label{jmu formula2}
J^\mu
= -2\left(1 + \alpha\right) \eta^{\mu\nu}
  \!\left(A_\nu^{(2)} + B_\nu^{(2)}\right)
  - \left(1 - \alpha \right) \eta^{\mu\nu}
    \eta^{\sigma t}\partial_\sigma F^{(0)}_{t\nu},
\end{equation}
where $\eta_{\mu\nu}$ is the Minkowski metric on the boundary. The $\alpha$-dependent corrections in~(\ref{jmu formula2}) arise from the modified asymptotic geometry and from the Gauss-Bonnet correction to the counter-term. In the limit $\alpha \to 0$, equation~(\ref{jmu formula2}) reduces to the known result for pure Einstein gravity~\cite{Bu:2015ame}.

\section{Off-Shell Constitutive Relation from Bulk Maxwell Dynamics}
\label{section:three}
The goal of this section is to derive the constitutive relation for the boundary $U(1)$ current and to express the transport coefficient functions $\mathcal{D}$, $\sigma_e$, and $\sigma_m$ in terms of bulk data. Our strategy, in the spirit of~\cite{Bu:2014sia, Bu:2014ena, Bu:2015bwa, Bu:2015ika, Bu:2015ame}, is to construct an \emph{off-shell} current: we solve only the dynamical components of the bulk Maxwell equations, which already suffice to completely fix the transport coefficients. The single remaining constraint equation is subsequently shown to be equivalent to the boundary continuity equation, thereby putting the current on-shell.

In the radial gauge $A_r = 0$, the five bulk Maxwell equations~(\ref{Maxwell eq}) decompose naturally as
\begin{align}
\text{dynamical equations:} \qquad
& \mathrm{EQ}^{\mu}=0,
\label{eq:dynamical}
\\
\text{constraint equation:} \qquad
& \mathrm{EQ}^{r}=0.
\label{eq:constraint}
\end{align}
The four dynamical equations \eqref{eq:dynamical} govern the evolution of the gauge field in the radial direction from the horizon outward to the conformal boundary. Their solutions, once substituted into the holographic formula~(\ref{jmu formula2}), fully determine the transport coefficients. The constraint equation \eqref{eq:constraint} does not participate in this construction; it is instead verified a posteriori to reproduce the continuity equation $\partial_\mu J^\mu = 0$.

We begin by identifying the static, uniform solution. When $A_\mu^{(0)}$ and $\rho$ are constants, independent of the boundary coordinates, the dynamical equations $\mathrm{EQ}^\mu = 0$ admit the exact solution
\begin{equation}\label{static}
A_\mu = A_\mu^{(0)} + \frac{\rho}{2r^2}\,\delta_{\mu t}.
\end{equation}
Substituting this into the holographic current formula~(\ref{jmu formula2}) gives
\begin{equation}
J^t = (1+\alpha)\rho, \qquad J^i = 0,
\end{equation}
from which we identify $\rho_{\rm phys}=(1+\alpha)\rho$ as the equilibrium charge density of the boundary plasma. This is the purely static configuration: a uniformly charged medium at rest with no external fields ($F_{\mu\nu}^{(0)} = 0$) and no dissipation.

To study the hydrodynamics of the $U(1)$ current, we follow the fluid/gravity philosophy~\cite{Bhattacharyya:2007vjd, Bu:2015ame} and promote the constant parameters to slowly varying functions of the boundary coordinates,
\begin{equation}
A_\mu^{(0)} \;\to\; A_\mu^{(0)}(x_\alpha), \qquad \rho \;\to\; \rho(x_\alpha).
\end{equation}
Once the parameters acquire spacetime dependence, the seed solution~(\ref{static}) ceases to satisfy the Maxwell equations. A bulk correction field $a_\mu(r, x_\alpha)$ must be introduced to restore consistency. The full ansatz is therefore
\begin{equation}\label{ansatz}
A_\mu(r, x_\alpha)
= A_\mu^{(0)}(x_\alpha) + \frac{\rho(x_\alpha)}{2r^2}\,\delta_{\mu t}
  + a_\mu(r, x_\alpha),
\end{equation}
where $a_\mu$ encodes all gradient corrections driven by the inhomogeneity of $A_\mu^{(0)}(x_\alpha)$ and $\rho(x_\alpha)$.

The following three conditions are imposed on the correction field $a_\mu$, together uniquely determining it.

\paragraph{i. Asymptotic condition.} Since $A_\mu^{(0)}(x_\alpha)$ is identified as the external electromagnetic potential that sources the current, the correction $a_\mu$ must not contribute to the leading boundary data. This requires
\begin{equation}\label{AdS constraint}
a_\mu(r \to \infty,\, x_\alpha) \;\longrightarrow\; 0.
\end{equation}

\paragraph{ii. Landau frame.}
We define the fluid velocity frame by the condition that the charge density as measured in the fluid rest frame is $(1+\alpha)\rho(x_\alpha)$. This is the Landau frame convention,\footnote{  In fluid mechanics, the stress tensor is decomposed as   $T^{\mu\nu} = T^{\mu\nu}_{\text{perfect}} + T^{\mu\nu}_{\text{diss}}$, with the   Landau frame defined by $u_\mu\,T^{\mu\nu}_{\text{diss}} = 0$ at every order~\cite{Bhattacharyya:2007vjd}.}
\begin{equation}\label{landau frame}
J^\mu u_\mu = -(1+\alpha)\rho(x_\alpha).
\end{equation}
As we will see, this condition translates into algebraic constraints on certain pre-asymptotic coefficients of $a_\mu$.

\paragraph{iii. Regularity at the horizon.} Working in ingoing Eddington-Finkelstein coordinates, the physical boundary condition at the black brane horizon is regularity: $a_\mu(r, x_\alpha)$ must be smooth for all $r \in [r_\text{h}, \infty)$. This is the bulk implementation of the in-falling wave condition~\cite{Son:2002sd}, which encodes the dissipative nature of the horizon. Together with the asymptotic condition~(\ref{AdS constraint}), regularity is sufficient to uniquely fix the correction field.

Substituting the ansatz~(\ref{ansatz}) into the dynamical Maxwell equations~(\ref{Maxwell eq}), the equations for $a_\mu$ are
\begin{equation}\label{eq_a}
\begin{split}
\mathrm{EQ}^t = 0 \;\Rightarrow\; 
0 =\;& (1+\alpha)\bigl(r^2\partial_r^2 a_t + 3r\,\partial_r a_t\bigr)
     + \partial_r\partial_k a_k, \\[4pt]
\mathrm{EQ}^i = 0 \;\Rightarrow\; 
0 =\;& \bigl[(r^9 - r^5) + \alpha(r^9 + r^5 + 2r)\bigr]\partial_r^2 a_i
     + \bigl[(3r^8 + r^4) + \alpha(3r^8 - r^4 - 10)\bigr]\partial_r a_i \\
    &+ 2r^7\partial_r\partial_t a_i
     - r^7\partial_r\partial_i a_t
     + r^6(\partial_t a_i - \partial_i a_t)
     + (1-\alpha)r^5(\partial^2 a_i - \partial_i\partial_k a_k) \\
    &+ \tfrac{1}{2}r^4\,\partial_i\rho
     + r^6\bigl(\partial_t A_i^{(0)} - \partial_i A_t^{(0)}\bigr)
     + (1-\alpha)r^5\bigl(\partial^2 A_i^{(0)} - \partial_i\partial_k A_k^{(0)}\bigr).
\end{split}
\end{equation}
The inhomogeneous terms on the right-hand side are sourced entirely by the boundary fields $A_\mu^{(0)}(x_\alpha)$ and $\rho(x_\alpha)$. It follows that $a_\mu$ is a linear functional of these two fields, a structure that we now exploit systematically.

Since $a_\mu$ is linear in $A_\mu^{(0)}$ and $\rho$, it admits a natural decomposition onto the basis of scalar and vector structures that can be formed from these fields and their derivatives,
\begin{equation}\label{basis}
\begin{split}
a_t &= S_1 A_t^{(0)} + S_2\,\partial_k A_k^{(0)} + S_3\,\rho, \\
a_i &= V_1 A_i^{(0)} + V_2\,\partial_i A_t^{(0)}
      + V_3\,\partial_i\partial_k A_k^{(0)} + V_4\,\partial_i\rho.
\end{split}
\end{equation}
The decomposition coefficients $S_i$ and $V_i$ are functions of the radial coordinate $r$, and in the momentum space representation, obtained via $\partial_t \to -i\omega$, $\partial_i \to iq_i$, they become functions of $(r, \omega, q^2)$. Inserting~(\ref{basis}) into~(\ref{eq_a}) and projecting onto each independent tensor structure, the coupled system of PDEs reduces to three decoupled sectors of second-order ODEs in the radial coordinate.

\paragraph{Sector I:~$\{S_1, V_2\}$}
\begin{equation}\label{S1V2}
\left\{
\begin{aligned}
0 =\;& (1+\alpha)r^2\partial_r^2 S_1
      + (1+\alpha)\,3r\,\partial_r S_1
      - q^2\partial_r V_2, \\
0 =\;& \bigl[(r^9 - r^5) + \alpha(r^9 + r^5 + 2r)\bigr]\partial_r^2 V_2
     + \bigl[(3r^8 + r^4) + \alpha(3r^8 - r^4 - 10)\bigr]\partial_r V_2 \\
    &- 2i\omega r^7\partial_r V_2
     - i\omega r^6 V_2
     - r^7\partial_r S_1
     - r^6 S_1
     - r^6.
\end{aligned}
\right.
\end{equation}

\paragraph{Sector II:~$\{S_2, V_1, V_3\}$}
\begin{equation}\label{S2V1V3}
\left\{
\begin{aligned}
0 =\;& (1+\alpha)r^2\partial_r^2 S_2
      + (1+\alpha)\,3r\,\partial_r S_2
      + \partial_r V_1
      - q^2\partial_r V_3, \\
0 =\;& \bigl[(r^9 - r^5) + \alpha(r^9 + r^5 + 2r)\bigr]\partial_r^2 V_1
     + \bigl[(3r^8 + r^4) + \alpha(3r^8 - r^4 - 10)\bigr]\partial_r V_1 \\
    &- 2i\omega r^7\partial_r V_1
     - i\omega r^6 V_1
     - (1-\alpha)q^2 r^5 V_1
     - i\omega r^6
     - (1-\alpha)q^2 r^5, \\
0 =\;& \bigl[(r^9 - r^5) + \alpha(r^9 + r^5 + 2r)\bigr]\partial_r^2 V_3
     + \bigl[(3r^8 + r^4) + \alpha(3r^8 - r^4 - 10)\bigr]\partial_r V_3 \\
    &- 2i\omega r^7\partial_r V_3
     - i\omega r^6 V_3
     - r^7\partial_r S_2
     - r^6 S_2
     - (1-\alpha)r^5 V_1
     - (1-\alpha)r^5.
\end{aligned}
\right.
\end{equation}

\paragraph{Sector III:~$\{S_3, V_4\}$}
\begin{equation}\label{S3V4}
\left\{
\begin{aligned}
0 =\;& (1+\alpha)r^2\partial_r^2 S_3
      + (1+\alpha)\,3r\,\partial_r S_3
      - q^2\partial_r V_4, \\
0 =\;& \bigl[(r^9 - r^5) + \alpha(r^9 + r^5 + 2r)\bigr]\partial_r^2 V_4
     + \bigl[(3r^8 + r^4) + \alpha(3r^8 - r^4 - 10)\bigr]\partial_r V_4 \\
    &- 2i\omega r^7\partial_r V_4
     - i\omega r^6 V_4
     - r^7\partial_r S_3
     - r^6 S_3
     + \tfrac{1}{2}r^4.
\end{aligned}
\right.
\end{equation}
The decoupling of Sectors I--III is a direct consequence of the tensor structure of the source terms in~(\ref{eq_a}), and persists in the presence of Gauss-Bonnet corrections. The $\alpha$-dependent modifications appear through two channels: corrections to the radial kinetic operators (the coefficients of $\partial_r^2$ and $\partial_r$ terms), and corrections to the spatial momentum coupling in Sector II. The boundary conditions for $S_i$ and $V_i$ inherit directly from those on $a_\mu$. The asymptotic condition~(\ref{AdS constraint}) imposes
\begin{equation}\label{Si Vi bcs}
S_i \;\longrightarrow\; 0, \qquad V_i \;\longrightarrow\; 0
\qquad \text{as } r \to \infty,
\end{equation}
while regularity of $a_\mu$ at the horizon is equivalent to regularity of each $S_i$ and $V_i$ there. The Landau frame condition imposes additional pre-asymptotic constraints on $S_i$, as we detail below.

The boundary current~(\ref{jmu formula2}) is determined by the subleading coefficients of $A_\mu$ near $r = \infty$, so we need the near-boundary behaviour of each $S_i$ and $V_i$. Using the ODE systems~(\ref{S1V2},~\ref{S2V1V3},~\ref{S3V4}) together with the boundary condition~(\ref{Si Vi bcs}), one finds that the characteristic exponents at $r = \infty$ are $0$ and $-2$ (relative to the leading $1/r$ fall-off), with logarithmic branches arising because the difference of the two characteristic indices is an integer. The resulting asymptotic expansions are
\begin{equation}\label{boundary behavior}
\begin{split}
S_1 &\longrightarrow
  \frac{s_1(\omega, q^2)}{r^2}
  - (1-2\alpha)\frac{q^2\log r}{2r^2}
  + \mathcal{O}\!\left(\frac{\log r}{r^3}\right), \\[3pt]
S_2 &\longrightarrow
  \frac{s_2(\omega, q^2)}{r^2}
  + (1-2\alpha)\frac{i\omega\log r}{2r^2}
  + \mathcal{O}\!\left(\frac{\log r}{r^3}\right), \\[3pt]
S_3 &\longrightarrow
  \frac{s_3(\omega, q^2)}{r^2}
  + \mathcal{O}\!\left(\frac{1}{r^3}\right), \\[3pt]
V_1 &\longrightarrow
  -(1-\alpha)\frac{i\omega}{r}
  + \frac{v_1(\omega, q^2)}{r^2}
  + \frac{1}{2}(1-2\alpha)(\omega^2 - q^2)\frac{\log r}{r^2}
  + \mathcal{O}\!\left(\frac{\log r}{r^3}\right), \\[3pt]
V_2 &\longrightarrow
  -(1-\alpha)\frac{1}{r}
  + \frac{v_2(\omega, q^2)}{r^2}
  - (1-2\alpha)\frac{i\omega\log r}{2r^2}
  + \mathcal{O}\!\left(\frac{\log r}{r^3}\right), \\[3pt]
V_3 &\longrightarrow
  \frac{v_3(\omega, q^2)}{r^2}
  - (1-2\alpha)\frac{\log r}{2r^2}
  + \mathcal{O}\!\left(\frac{\log r}{r^3}\right), \\[3pt]
V_4 &\longrightarrow
  \frac{v_4(\omega, q^2)}{r^2}
  + \mathcal{O}\!\left(\frac{1}{r^3}\right).
\end{split}
\end{equation}
The logarithmic coefficients are fixed by the ODE systems and the boundary conditions, while the finite coefficients $s_i(\omega, q^2)$ and $v_i(\omega, q^2)$ are undetermined by pre-asymptotic analysis alone and must be found by integrating outward from the horizon. They carry the full dynamical information about the  transport.

Substituting~(\ref{boundary behavior}) into the holographic formula~(\ref{jmu formula2}), the boundary current becomes
\begin{equation}\label{jmu original}
\begin{split}
J^t =\;& (1+\alpha)\bigl[\rho + 2s_1 A_t^{(0)}
         + 2s_2\,\partial_k A_k^{(0)} + 2s_3\,\rho\bigr]
        + \tfrac{1}{2}(1-\alpha)\bigl[q^2 A_t^{(0)}
        - i\omega\,\partial_k A_k^{(0)}\bigr], \\[4pt]
J^i =\;& -\bigl[2(1+\alpha)v_1
          + \tfrac{1}{2}(1-\alpha)(\omega^2 + q^2)\bigr]A_i^{(0)}
         - \bigl[2(1+\alpha)v_2
         - \tfrac{1}{2}(1-\alpha)\,i\omega\bigr]\partial_i A_t^{(0)} \\
        &- \bigl[2(1+\alpha)v_3
          + \tfrac{1}{2}(1-\alpha)\bigr]\partial_i\partial_k A_k^{(0)}
         - (1+\alpha)\,2v_4\,\partial_i\rho.
\end{split}
\end{equation}

\paragraph{Landau frame and gauge invariance constraints.} Imposing the Landau frame convention~(\ref{landau frame}) on the time component of~(\ref{jmu original}) fixes three of the pre-asymptotic coefficients algebraically,
\begin{equation}\label{lf}
s_1 = -\frac{1-2\alpha}{4}\,q^2,
\qquad
s_2 = \frac{1-2\alpha}{4}\,i\omega,
\qquad
s_3 = 0.
\end{equation}
A further set of constraints arises from demanding that the boundary current be invariant under bulk-induced gauge transformations of the external potential,
\begin{equation}\label{bdry gauge}
J^\mu \to J^\mu \qquad \text{under} \qquad A_\mu^{(0)}(x_\alpha) \to A_\mu^{(0)}(x_\alpha) + \partial_\mu\varphi(x_\alpha).
\end{equation}
This gauge invariance requires the combinations $X = i\omega S_1 + q^2 S_2$ and $Y = i\omega V_2 + q^2 V_3 - V_1$ to vanish identically as functions of $r$,
\begin{equation}\label{constraint relations1}
i\omega S_1 + q^2 S_2 = 0, \qquad i\omega V_2 + q^2 V_3 - V_1 = 0.
\end{equation}
The first of these is automatically satisfied by the Landau frame choice~(\ref{lf}). The second gives a relation among the pre-asymptotic coefficients,
\begin{equation}\label{constraint relations2}
i\omega\,v_2 + q^2 v_3 - v_1 = 0,
\end{equation}
which we use to eliminate $v_1$ in favour of $v_2$ and $v_3$, reducing the number of independent unknowns.

\paragraph{Gauge-invariant current and transport coefficients.} Utilizing~(\ref{constraint relations2}) to eliminate $v_1$, and expressing the result through the gauge-invariant field strengths $F_{ti}^{(0)}$ and $\partial_k F_{ki}^{(0)}$, the spatial current takes the manifestly 
gauge-invariant form
\begin{equation}
J^i = -(1+\alpha)\,2v_4\,\partial_i\rho
      + \bigl[2(1+\alpha)v_2 - \tfrac{1}{2}i\omega(1-\alpha)\bigr]F_{ti}^{(0)}
      + \bigl[2(1+\alpha)v_3 + \tfrac{1}{2}(1-\alpha)\bigr]\partial_k F_{ki}^{(0)}.
\end{equation}
Comparing with the general constitutive relation
\begin{equation}
J^t=\rho_{\rm phys},
\qquad
J^i
=-\mathcal{D}(\omega,q^2)\,\partial_i\rho_{\rm phys}
+\sigma_e(\omega,q^2)\,F_{ti}^{(0)}
+\sigma_m(\omega,q^2)\,\partial_k F_{ki}^{(0)} \, ,
\label{general_const_reltn}
\end{equation}
we read off the three transport coefficient functions,
\begin{equation}\label{Dv}
\mathcal{D} = 2v_4,
\qquad
\sigma_e = 2(1+\alpha)v_2 - \tfrac{1}{2}i\omega(1-\alpha),
\qquad
\sigma_m = 2(1+\alpha)v_3 + \tfrac{1}{2}(1-\alpha).
\end{equation}
The problem of determining the all-order transport coefficient functions has thus been reduced to computing the three pre-asymptotic boundary coefficients $v_2$, $v_3$, and $v_4$, the finite terms in the near-boundary expansion~(\ref{boundary behavior}). These cannot be extracted from near-boundary analysis alone; they require integrating the ODE systems~(\ref{S1V2},~\ref{S2V1V3},~\ref{S3V4}) across the full bulk from the horizon to the boundary. This is the content of the next section, where we carry out this integration both analytically in the hydrodynamic regime and numerically for generic momenta.

\section{Hydrodynamic expansion: analytic Gauss-Bonnet corrections}\label{section:four}
We now turn to the long-wavelength regime, where the transport data can be obtained analytically. Since the constitutive relation derived in the previous section is already fixed
in terms of the near-boundary coefficients $v_i$, the remaining task is to solve the radial equations perturbatively for small frequency and momentum, and then read off the corresponding expansion of the transport functions.

The hydrodynamic regime is defined by
\[
\omega \ll 1, \qquad q \ll 1,
\]
with all quantities measured in units of $\pi T$. In this regime it is convenient to introduce a formal bookkeeping parameter $\lambda$ which keeps track of the order in the gradient expansion,
\begin{equation}
\omega \to \lambda \omega, \qquad \vec{q} \to \lambda \vec{q}.
\label{hydro_lambda}
\end{equation}

The radial coefficient functions appearing in the decomposition of the bulk correction field are then expanded as formal series in $\lambda$,
\begin{equation}
S_i(r,\omega,\vec{q})=\sum_{n=0}^{\infty}\lambda^n S_i^{(n)}(r,\omega,\vec{q}),
\qquad
V_i(r,\omega,\vec{q})=\sum_{n=0}^{\infty}\lambda^n V_i^{(n)}(r,\omega,\vec{q}).
\label{hydro_series}
\end{equation}
Substituting \eqref{hydro_series} into the ODE system yields an iterative set of radial equations, to be solved order by order in $\lambda$. At each order, the integration constants are fixed by regularity at the horizon, vanishing source deformation at the boundary, and the Landau-frame condition. The explicit perturbative radial solutions are given in appendix~\ref{appendixA}.

From their near-boundary behaviour one reads off the coefficients entering the constitutive relation and we find
\begin{equation}
\begin{aligned}
v_1 &=
\frac{1}{2}(1-2\alpha)i\omega
-\frac{1}{4}\Big[q^2+\omega^2(1+\log2-6\alpha)\Big]
+\cdots,
\\[2mm]
v_2 &=
\frac{1}{2}(1-2\alpha)
+\frac{1}{4}i\omega(1+\log2-6\alpha)
+\cdots,
\\[2mm]
v_3 &= -\frac{1}{4}+\cdots,
\\[2mm]
v_4 &= \frac{1}{4}
+\frac{i\omega}{48}\Big[3\pi(1+2\alpha)-16\alpha\Big]
+\cdots .
\end{aligned}
\label{hydro_vi}
\end{equation}
These coefficients determine the hydrodynamic expansion of the transport functions through the identifications obtained in section \ref{section:three}. Substituting \eqref{hydro_vi} in \eqref{Dv}, we obtain
\begin{align}
\mathcal{D}(\omega,q^2)
&=
\frac{1}{2}
+\frac{i\omega}{24}\Big[3\pi(1+2\alpha)-16\alpha\Big]
+\cdots, \label{hydro_transport_diff}
\\[2mm]
\sigma_e(\omega,q^2)
&=
(1-\alpha)
+\frac{i\omega}{2}\Big[\log2+\alpha(\log2-4)\Big]
+\cdots, \label{hydro_transport_sigmae}
\\[2mm]
\sigma_m(\omega,q^2)
&=
-\alpha+\cdots \label{hydro_transport_sigmam}.
\end{align}

Several features are worth noting. The diffusion function remains finite in the Gauss-Bonnet background and continues to admit a regular derivative expansion around $\omega=q=0$, although its first frequency correction is shifted by $\alpha$. The electric conductivity is modified already at the level of the DC term, and the coefficient of the linear frequency correction also receives a nontrivial higher-curvature contribution. The magnetic conductivity is particularly interesting: while it does not contribute at leading order in the Einstein limit, in the present case a nonvanishing zeroth-order term is generated by the Gauss-Bonnet deformation. This is one of the clearest indications that the higher-derivative background changes the off-shell constitutive structure itself, rather than merely perturbing a single transport coefficient.

\section{All-order transport functions and memory kernel}\label{section:five}
Away from the hydrodynamic regime, the transport functions are determined by solving the radial equations for generic values of $(\omega,q^2)$. This is a boundary value problem for a coupled system of linear second-order ordinary differential equations. We solve it by a spectral collocation method, following the strategy employed in Ref.~\cite{Bu:2016oba}. The basic idea is to replace the continuous radial problem by a finite-dimensional linear algebra problem. After mapping the radial domain to a finite interval, the unknown functions are represented by their values on a Chebyshev collocation grid. Derivatives with respect to the radial coordinate are then approximated by differentiation matrices constructed from polynomial interpolation. In this way the differential equations are converted into a linear algebraic system for the values of the functions on the grid points, while the horizon regularity and boundary conditions are imposed directly at the endpoints of the grid. This approach is particularly well suited to the present problem, since it treats the horizon and the conformal boundary on equal footing and avoids the iterative tuning of initial data characteristic of shooting methods.

\subsection{Diffusion function}

\begin{figure}[t!]
\centering
\includegraphics[width=\linewidth]{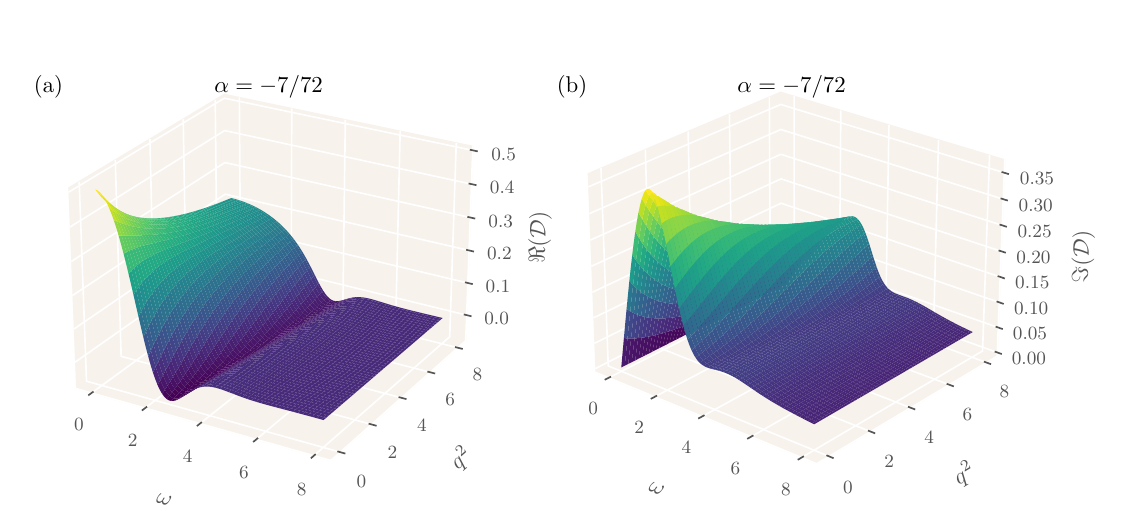}
\caption{(a) Real and (b) imaginary parts of the diffusion function $\mathcal{\tilde  D}(\omega,q^2)$ shown as functions of $\omega$ and $q^2$ for a fixed $\alpha$. }
\label{figure1}
\end{figure}

\begin{figure}[t!]
\centering

\includegraphics[width=\linewidth]{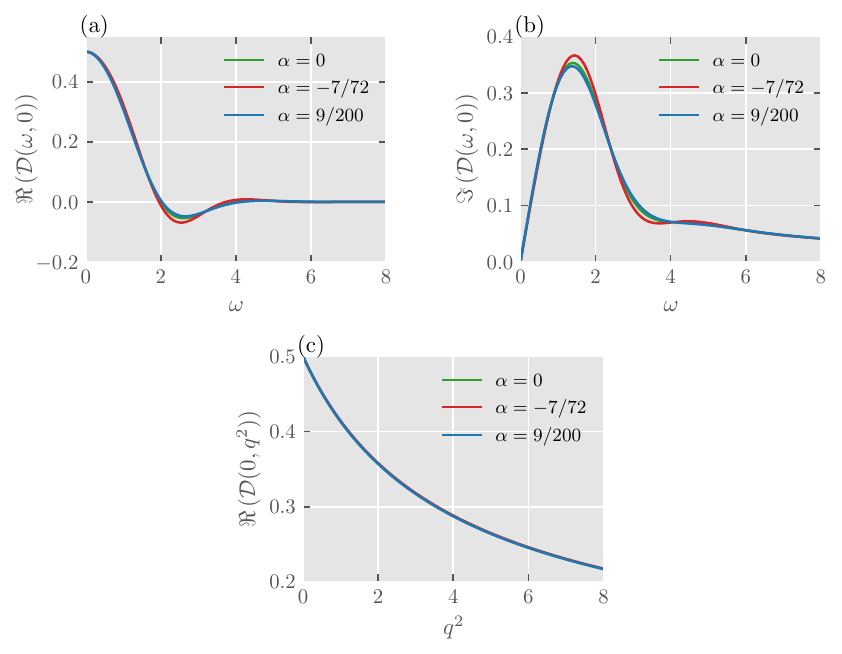}
\caption{Two-dimensional profiles of the diffusion function $\mathcal{D}(\omega, q^2)$ for different values of $\alpha$. (a) real and (b) imaginary parts as functions of $\omega$ at $q=0$. (c) real part as a function of $q^2$ at $\omega=0$. }
\label{figure2}
\end{figure}

We begin with the diffusion function $\mathcal D(\omega,q^2)$. The numerical results show that the dependence on $q^2$ remains relatively mild, whereas the frequency dependence is considerably richer. The full $(\omega,q^2)$ behaviour is displayed in Fig.~\ref{figure1}. Both the real and imaginary parts develop damped oscillations as functions of $\omega$, and their amplitudes decrease at large frequency and momentum. This pattern is characteristic of an all-order resummed transport function and reflects the influence of the set of quasi-normal modes in the bulk description~\cite{Bu:2015ame}. At sufficiently large $(\omega,q^2)$, $\mathcal D$ tends to zero, indicating that the resummed constitutive relation suppresses instantaneous diffusive response outside the hydrodynamic regime. This behaviour is consistent with the causal interpretation of the effective theory, which will be examined more directly in the time domain below.

The two-dimensional sections in Fig.~\ref{figure2} make these features more transparent. At $\omega=0$, the diffusion function varies smoothly with $q^2$, and the imaginary part vanishes, as expected. By contrast, at $q=0$ both $ \Re \,(\mathcal D)$ and $ \Im\,(\mathcal D)$ show an oscillatory dependence on $\omega$ before decaying at larger frequency. Thus the nontrivial structure of the resummed diffusion function is controlled primarily by temporal, rather than spatial, gradients. In this sense, the diffusion channel remains quasi-local in momentum space while retaining a pronounced nonlocality in time. This is the same qualitative mechanism that underlies the appearance of a memory kernel in the constitutive relation. 

From the point of view of the Gauss-Bonnet deformation, the main effect is not to alter the overall pattern of the diffusion function, but to shift it quantitatively. The hydrodynamic expansion already showed that the leading diffusion constant is unchanged at this order, whereas the first frequency correction acquires an explicit $\alpha$-dependence~\eqref{hydro_transport_diff}. The numerical results indicate that this persists beyond the strict hydrodynamic limit: the Gauss-Bonnet coupling modifies the location and profile of the oscillatory structure, but does not remove the characteristic damping or the large-momentum falloff. Physically, this means that the higher-curvature correction renormalises the relaxation pattern of charge transport through its effect on the bulk quasi-normal spectrum, while leaving intact the basic resummed diffusive behaviour of the Einstein case. Put differently, $\alpha$ deforms the relaxation time scales rather than the existence of the diffusive channel itself.

It is also instructive to compare $\mathcal D$ with the all-order shear-viscosity function obtained previously in Einstein and Einstein-Gauss-Bonnet gravity~\cite{Bu:2014ena, Bu:2015bwa}. The two functions are
not proportional, either in the derivative expansion or at generic momenta, yet their overall behaviour is qualitatively similar. This suggests that the damped oscillatory structure is not specific to the charge sector, but is instead a broader feature of resummed dissipative transport in holographic plasmas.

\subsection{Memory function and causality}
\begin{figure}[t!]
\centering
\includegraphics[width=\linewidth]{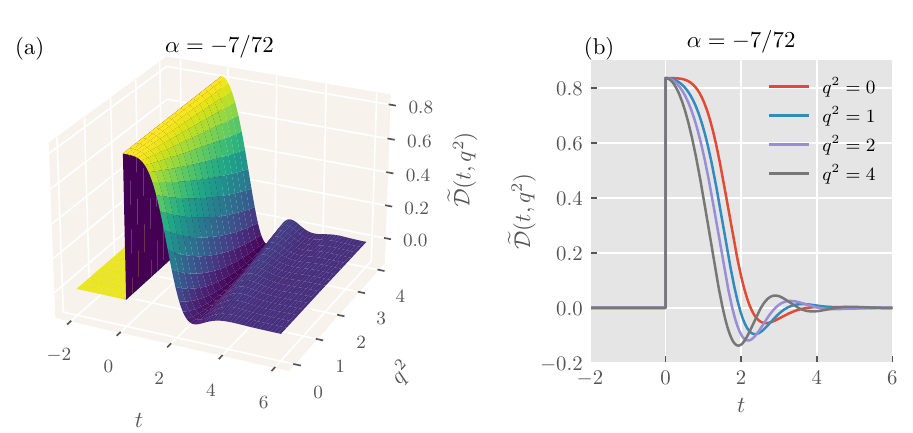}
\caption{Memory function $\mathcal{D}(t,q^2)$. (a) 3D representation as a function of $t$ and $q^2$ for fixed $\alpha$; (b) 2D profiles as functions of $t$ for several values of $q^2$ at fixed $\alpha$.}
\label{figuremem1}
\end{figure}

\begin{figure}[t!]
\centering
\includegraphics[width=\linewidth]{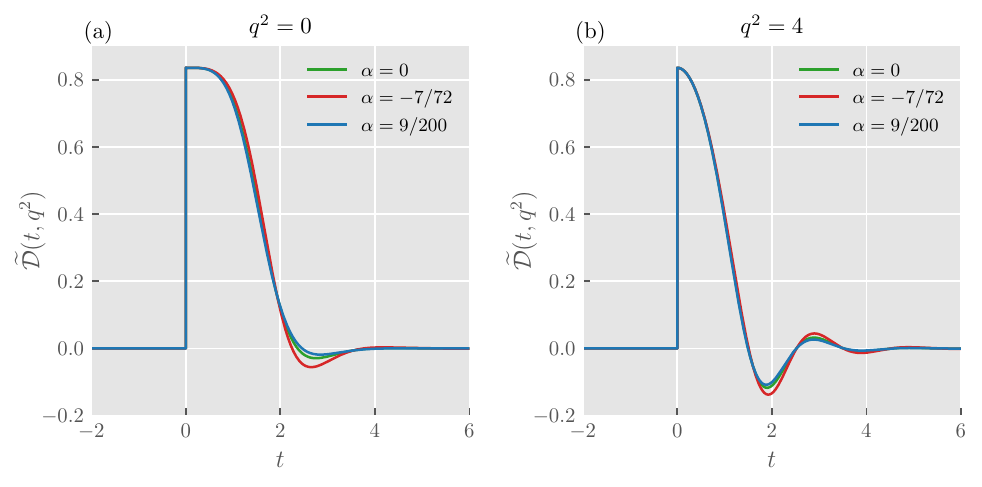}
\caption{2D profiles of the memory function $\mathcal{\tilde D}(t,q^2)$ for several values of $\alpha$: (a) $q^2=0$ and (b) $q^2=4$.}
\label{figuremem2}
\end{figure}

A more direct test of causality is obtained by Fourier transforming the diffusion function to the time domain. We define the memory kernel by
\begin{equation}\label{eq:memory_kernel_new}
\widetilde{\mathcal D}(t,q^2)=\int_{-\infty}^{\infty}\frac{d\omega}{2\pi}\,\mathcal D(\omega,q^2)e^{-i\omega t}.
\end{equation}
In the constitutive relation, the diffusion part of the current may then be written as a temporal convolution,
\begin{equation}\label{eq:memory_constitutive_new}
J_i^{\rm diff}(t,\vec x)
=-\int_{-\infty}^{\infty}dt'\,\widetilde{\mathcal D}(t-t',q^2)\,\partial_i\rho(t',\vec x).
\end{equation}
Our numerical results, Figs.~\ref{figuremem1} and \ref{figuremem2}, show that $\widetilde{\mathcal D}(t,q^2)$ has no support for $t<0$ up to numerical noise. In this sense the all-order resummation converts the formally acausal Fick-type truncation into a causal constitutive law with finite memory. For the present problem, the Gauss-Bonnet deformation changes the detailed shape of the kernel but does not spoil this causal structure.

Figs.~\ref{figuremem1} and \ref{figuremem2} illustrate how spatial momentum and the Gauss–Bonnet coupling modify the temporal profile of the memory function. At fixed \(\alpha\), increasing \(q^2\) leaves the initial amplitude nearly unchanged but shifts the first zero crossing to earlier times and produces a deeper negative minimum, followed by more pronounced damped oscillations. The dependence on \(\alpha\) is comparatively modest: negative \(\alpha\) deepens the negative minimum and enhances the subsequent positive excursion, whereas positive \(\alpha\) suppresses these features relative to the Einstein case. This effect is visible at both \(q^2=0\) and \(q^2=4\). In all cases, the kernel decays towards zero at late times and remains vanishing for \(t<0\) within numerical accuracy. Thus, spatial momentum and higher-curvature corrections reshape the transient response while preserving its retarded character.

\subsection{Electric and magnetic conductivities}

\begin{figure}[t!]
\centering
\includegraphics[width=\linewidth]{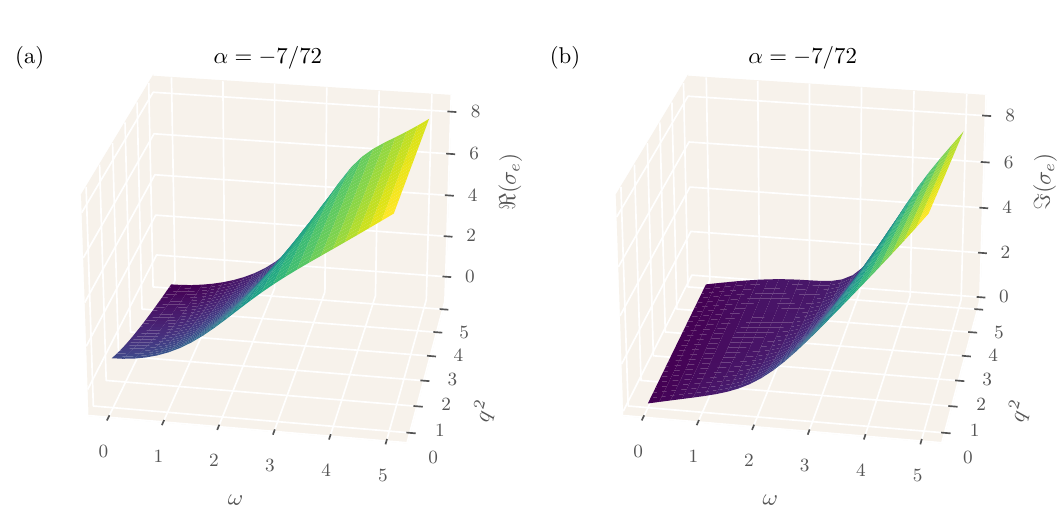}
\caption{(a) Real and (b) imaginary parts of the electric conductivity $\sigma_e(\omega,q^2)$ shown as functions of $\omega$ and $q^2$ for a fixed $\alpha$ value.}
\label{figure4}
\end{figure}
\begin{figure}[htbp]
\centering
\includegraphics[width=\linewidth]{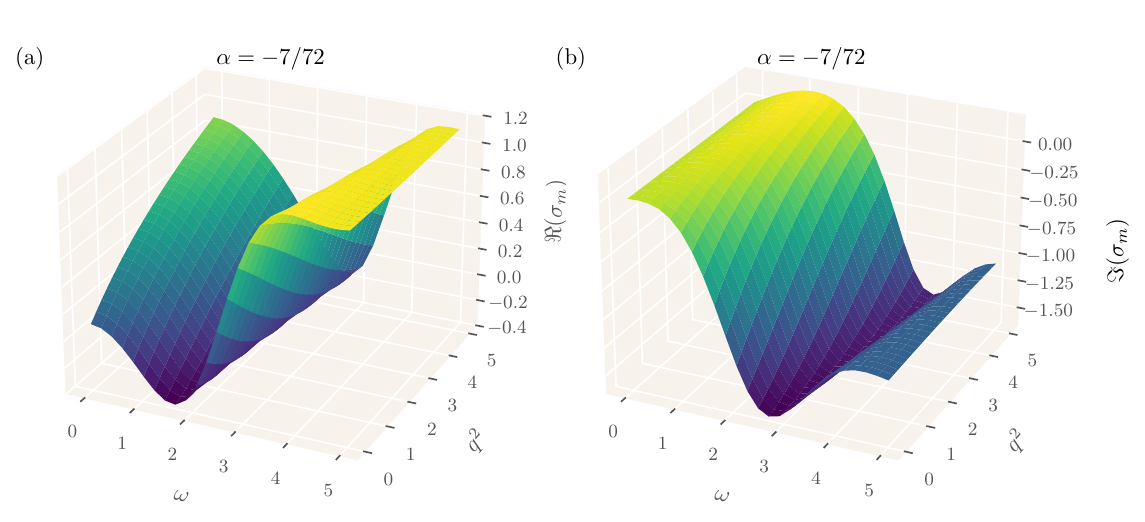}
\caption{(a) Real and (b) imaginary parts of the magnetic conductivity $\sigma_m(\omega,q^2)$ shown as functions of $\omega$ and $q^2$ for a fixed $\alpha$ value.}
\label{figure5}
\end{figure}

\begin{figure}[t!]
\centering
\includegraphics[width=\linewidth]{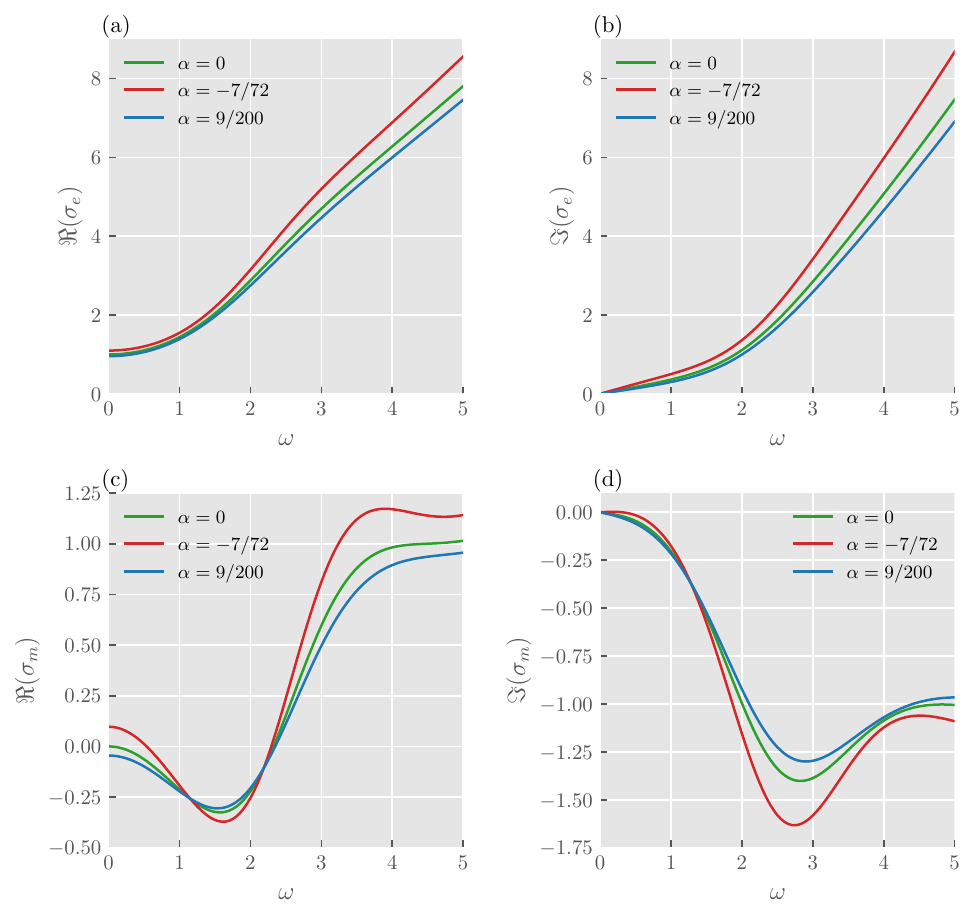}
\caption{Frequency dependence of the conductivities at $q=0$ for several $\alpha$ values. (a) real and (b) imaginary parts of $\sigma_e(\omega,0)$. (c) real and (d) imaginary parts of $\sigma_m(\omega,0)$.}
\label{figure6}
\end{figure}

We next consider the all-order conductivities $\sigma_e(\omega,q^2)$ and $\sigma_m(\omega,q^2)$. As in the diffusion channel, the numerical solutions show a relatively mild dependence on $q^2$ and a substantially richer dependence on $\omega$. At zero spatial momentum, the conductivity provides several useful checks of the numerical procedure. In the Einstein limit, $\alpha=0$, the numerical result for $\sigma_e(\omega,0)$ reproduces the known analytic expression in terms of digamma functions over the full frequency range~\cite{Bu:2015ame, Horowitz:2008bn, Myers:2007we}. For nonzero Gauss--Bonnet coupling, the zero-frequency limit agrees with the hydrodynamic result \eqref{hydro_transport_sigmae}, $\sigma_e(0,0)=1-\alpha+\mathcal{O}(\alpha^2)$. In addition, the small-frequency behavior of \eqref{hydro_transport_sigmae} satisfies
\begin{equation}
\frac{\Im\, \left[ \sigma_e(\omega,0) \right]}{\omega}
\xrightarrow{\omega\to0}
\frac{1}{2}\left[\log 2+\alpha(\log 2-4)\right],
\end{equation}
providing an independent check of the Gauss--Bonnet correction to the conductivity. 

The full $(\omega,q^2)$ dependence of $\sigma_e$ and $\sigma_m$ is displayed in Figs.~\ref{figure4} and \ref{figure5}. In both cases the variation along the $q^2$ direction is weak compared to the variation with $\omega$, indicating that the conductivity sector, like the diffusion function, is governed mainly by temporal gradients. The electric conductivity grows monotonically with frequency, with both its real and imaginary parts increasing as $\omega$ is raised. By contrast, the magnetic conductivity shows a more structured profile: its real part develops a pronounced dip at intermediate frequency, while the imaginary part becomes negative and reaches a minimum before gradually flattening at larger $\omega$. This difference is natural from the constitutive relation~\eqref{general_const_reltn}. The coefficient $\sigma_e$ governs the direct response to the applied electric field, whereas $\sigma_m$ multiplies a derivative of the field strength and therefore probes a subleading, genuinely gradient-induced part of the response.

The Gauss-Bonnet coupling modifies these conductivities in a channel-dependent way. For $\sigma_e$, the effect is primarily quantitative. Figure~\ref{figure6} shows that, at $q=0$, negative $\alpha$ enhances both the real and imaginary parts, while positive $\alpha$ suppresses them. This behaviour is already anticipated by the hydrodynamic expansion~\eqref{hydro_transport_sigmae}, so the numerical solutions indicate that the same ordering with respect to $\alpha$ persists beyond the strict low-frequency regime. In physical terms, the higher-curvature deformation shifts the strength of the electric response without altering its overall frequency profile.

The effect on $\sigma_m$ is more distinctive. In the Einstein limit the magnetic conductivity vanishes at leading order, whereas the Gauss-Bonnet deformation generates a nonzero static contribution \eqref{hydro_transport_sigmam}. This is visible directly in Fig.~\ref{figure6}: at $\omega=0$, $\sigma_m$ starts from positive values for negative $\alpha$, vanishes for $\alpha=0$, and becomes negative for positive $\alpha$. As $\omega$ increases, the three curves retain the same qualitative shape but are shifted relative to one another, with negative $\alpha$ generally enhancing the response and positive $\alpha$ suppressing it. Thus the Gauss-Bonnet correction does not merely renormalise an existing conductivity. Rather, it activates and reshapes the higher-derivative magnetic response itself.

Taken together, Figs.~\ref{figure4}--\ref{figure6} show that the conductivity sector is sensitive to the Gauss-Bonnet deformation in a more differentiated manner than the diffusion channel. The electric conductivity preserves the basic structure familiar from Einstein gravity, but with shifted magnitude, while the magnetic conductivity provides a clearer signature of the higher-curvature background through the appearance of a nonvanishing low-frequency contribution and an $\alpha$-dependent distortion of its frequency profile.

\section{Conclusion}\label{section:six}

In this work, we investigated the effect of Gauss–Bonnet corrections on holographic \(U(1)\) charge transport within the all-order linearised fluid/gravity framework. The bulk description consists of a probe Maxwell field in a five-dimensional asymptotically AdS black-brane geometry, which isolates the charge sector from energy-momentum backreaction. Working to first order in the Gauss–Bonnet coupling, we obtained the renormalised boundary current and expressed its off-shell constitutive relation in terms of three independent transport functions: the diffusion function \(\mathcal{D}(\omega,q^2)\), the electric conductivity \(\sigma_e(\omega,q^2)\), and the magnetic conductivity \(\sigma_m(\omega,q^2)\). This construction retains the full derivative dependence of the linear response and allows higher-curvature effects to be studied beyond a finite hydrodynamic truncation.

The analytic hydrodynamic expansion reveals a distinct response to the Gauss–Bonnet deformation in each transport channel. In the normalization used throughout this work, the leading diffusion constant remains \(\mathcal{D}(0,0)=1/2\), while the first frequency correction acquires an explicit dependence on \(\alpha\). The electric conductivity is modified already in the DC limit, with \(\sigma_e(0,0)=1-\alpha+\mathcal{O}(\alpha^2)\), and its first frequency correction is shifted as well. A particularly notable result is the appearance of \(\sigma_m(0,0)=-\alpha+\mathcal{O}(\alpha^2)\), which vanishes in the Einstein limit. Since this coefficient multiplies a spatial derivative of the external field strength, it represents a leading contribution to the response to an inhomogeneous magnetic source. The higher-curvature deformation therefore affects different components of the current in ways that cannot be captured by a common rescaling of the Einstein-gravity transport data.

Beyond the hydrodynamic regime, we determined the transport functions using spectral collocation. Their dependence on frequency is substantially richer than their dependence on spatial momentum over the range explored. The diffusion function retains the damped oscillatory behaviour familiar from the Schwarzschild–AdS background and decreases at large frequency and momentum, while the Gauss–Bonnet coupling shifts its profile. The conductivities display a different pattern. Negative \(\alpha\) enhances the electric response, whereas positive \(\alpha\) suppresses it in the numerical results. The magnetic conductivity exhibits a more structured frequency dependence, with coupling-dependent extrema and a static contribution whose sign is opposite to that of \(\alpha\). The Einstein limit and the analytic low-frequency expansion provide complementary checks of the numerical calculation.

The time-domain description gives a direct interpretation of the resummed diffusion function. Its inverse Fourier transform defines a memory kernel through which the diffusive current depends on the previous evolution of the charge density. For the couplings and momenta examined, the kernel has no support at negative times within numerical accuracy. The Gauss–Bonnet deformation changes the temporal response while preserving this retarded structure. These results support the persistence of the memory-based constitutive description when higher-curvature corrections are included.

Several extensions follow naturally from the present analysis. A direct study of the poles of the transport functions would clarify how the modified frequency profiles are related to shifts in the bulk quasinormal spectrum. Going beyond the probe approximation would allow charge diffusion to be studied together with energy-momentum transport at finite density. Extending the calculation beyond linear order in the Gauss–Bonnet coupling would also help establish the range of validity of the perturbative results. The present framework provides a basis for these investigations by determining how higher-curvature effects enter both the low-frequency transport coefficients and the resummed current response.

\appendix

\section{Perturbative solutions for $S_i$ and $V_i$}
\label{appendixA}

In this appendix we solve the radial equations \eqref{S1V2}, \eqref{S2V1V3}, and \eqref{S3V4} perturbatively in the hydrodynamic expansion parameter $\lambda$, introduced through
\[
\omega\rightarrow\lambda\omega,
\qquad
q\rightarrow\lambda q.
\]
The coefficient functions are expanded as
\begin{equation}
S_i(r,\omega,q^2)
 =
 \sum_{n=0}^{\infty}\lambda^n S_i^{(n)}(r,\omega,q^2),
\qquad
V_i(r,\omega,q^2)
 =
 \sum_{n=0}^{\infty}\lambda^n V_i^{(n)}(r,\omega,q^2).
\label{app:hydro-expansion}
\end{equation}
All expressions below are retained to linear order in the Gauss--Bonnet coupling $\alpha$.

\subsection*{Horizon-fixed perturbative formulation}

The horizon is located at
\begin{equation}
r_\text{h}=1-\alpha\equiv R.
\label{app:horizon}
\end{equation}
The radial kinetic operator appearing in the equations for the vector functions, \eqref{S1V2}, \eqref{S2V1V3}, and \eqref{S3V4}, can be written as
\begin{equation}
\partial_r\left[H(r)\partial_r X(r)\right],
\qquad
H(r)\equiv Rr^3f(r),
\label{app:H-definition}
\end{equation}
where
\begin{equation}
f(r)
 =
 1-\frac{1}{r^4}
 +2\alpha\left(1+\frac{1}{r^8}\right).
\end{equation}
To first order in $\alpha$,
\begin{equation}
H(r)
 =
 r^3-\frac{1}{r}
 +\alpha\left(
 r^3+\frac{1}{r}+\frac{2}{r^5}
 \right)
 +\mathcal{O}(\alpha^2).
\label{app:H-expanded}
\end{equation}
A direct expansion of the inverse radial operator at fixed $r$ gives
\begin{equation}
\frac{1}{H(r)}
 =
 \frac{r}{r^4-1}
 -
 \alpha
 \frac{r^8+r^4+2}
 {r^3(r^4-1)^2}
 +\mathcal{O}(\alpha^2).
\label{app:nonuniform-kernel}
\end{equation}
This expansion is not uniform near the horizon. Indeed, close to $r=1$,
\begin{equation}
\frac{1}{H(r)}
 \sim
 \frac{1}{4(r-1)}
 -
 \frac{\alpha}{4(r-1)^2}
 +\cdots,
\end{equation}
whereas the physical horizon is at $r_\text{h}=1-\alpha$. Since $|r_\text{h}-1|=\mathcal{O}(\alpha)$, the expansion \eqref{app:nonuniform-kernel} fails in the region in which horizon regularity must be imposed.

To perform the perturbative expansion consistently, we introduce the horizon-fixed radial coordinate
\begin{equation}
z\equiv\frac{r}{r_\text{h}}=\frac{r}{R},
\qquad
1\leq z<\infty.
\label{app:z-coordinate}
\end{equation}
In terms of $z$,
\begin{equation}
\begin{split}
f(Rz)
 &=
 1-\frac{1}{R^4z^4}
 +2\alpha\left(1+\frac{1}{R^8z^8}\right)
 \\
 &=
 \left(1-\frac{1}{z^4}\right)
 \left[
 1+2\alpha\left(1-\frac{1}{z^4}\right)
 \right]
 +\mathcal{O}(\alpha^2).
\end{split}
\label{app:f-fixed-horizon}
\end{equation}
Since $\partial_r=R^{-1}\partial_z$, one finds
\begin{equation}
H(r)\partial_r X
 =
 \mathcal{H}(z)\partial_z X,
\end{equation}
where
\begin{equation}
\mathcal{H}(z)
 =
 R^3z^3f(Rz)
 =
 z^3(1-z^{-4})
 \left[
 1-\alpha(1+2z^{-4})
 \right]
 +\mathcal{O}(\alpha^2).
\label{app:Hcal}
\end{equation}
Consequently,
\begin{equation}
\partial_r\left(H\partial_rX\right)
 =
 \frac{1}{R}
 \partial_z\left(\mathcal{H}\partial_zX\right).
\label{app:operator-transform}
\end{equation}

Near the physical horizon $z=1$,
\begin{equation}
\mathcal{H}(z)
 =
 4(1-3\alpha)(z-1)
 +\mathcal{O}\left((z-1)^2,\alpha^2\right).
\label{app:Hcal-horizon}
\end{equation}
Thus, if
\begin{equation}
\partial_z\left(\mathcal{H}X_z\right)
 =
 \mathcal{S}(z),
\end{equation}
the solution that is regular at $z=1$ and vanishes at the boundary is
\begin{equation}
X(z)
 =
 -\int_z^\infty
 \frac{d\xi}{\mathcal{H}(\xi)}
 \int_1^\xi
 \mathcal{S}(\eta)\,d\eta.
\label{app:general-regular-solution}
\end{equation}
The inner integral vanishes at $\xi=1$ and therefore cancels the simple zero of $\mathcal{H}(\xi)$. This representation is manifestly regular at the physical horizon.

For later convenience, we also define
\begin{equation}
\beta\equiv\frac{R}{1+\alpha}
 =
 1-2\alpha+\mathcal{O}(\alpha^2).
\label{app:beta}
\end{equation}

\paragraph{I: $\{S_1,V_2\}$.}

At the first two orders,
\begin{equation}
S_1^{(0)}=S_1^{(1)}=0.
\label{app:S1-low-orders}
\end{equation}
The zeroth-order equation for $V_2$ is
\begin{equation}
\partial_r\left(H\partial_rV_2^{(0)}\right)=1.
\label{app:V20-equation-r}
\end{equation}
Regularity at $r=r_\text{h}$ fixes the first integration constant:
\begin{equation}
H(r)\partial_rV_2^{(0)}(r)=r-r_\text{h}.
\label{app:V20-flux}
\end{equation}
Imposing $V_2^{(0)}(\infty)=0$ gives
\begin{equation}
V_2^{(0)}(r)
 =
 -\int_r^\infty
 \frac{x-r_\text{h}}{H(x)}\,dx.
\label{app:V20-r}
\end{equation}
In the horizon-fixed coordinate this becomes
\begin{equation}
V_2^{(0)}(z)
 =
 -N\int_z^\infty
 \frac{\xi-1}{\mathcal{H}(\xi)}\,d\xi.
\label{app:V20-z}
\end{equation}
Since
\[
\mathcal{H}(z)
 =
 4(1-3\alpha)(z-1)+\cdots,
\]
the horizon derivative is finite:
\begin{equation}
\partial_zV_2^{(0)}(1)
 =
 \frac{R}{4(1-3\alpha)}.
\end{equation}
The near-boundary expansion is
\begin{equation}
V_2^{(0)}
 =
 -\frac{1-\alpha}{r}
 +\frac{1-2\alpha}{2r^2}
 +\mathcal{O}\left(\frac{1}{r^3}\right).
\label{app:V20-asymptotic}
\end{equation}

At first order in the hydrodynamic expansion,
\begin{equation}
\partial_z
\left(
\mathcal{H}\partial_zV_2^{(1)}
\right)
 =
 R i\omega
 \left(
 2z\partial_zV_2^{(0)}+V_2^{(0)}
 \right).
\label{app:V21-equation}
\end{equation}
The regular solution is
\begin{equation}
V_2^{(1)}(z)
 =
 -Ni\omega
 \int_z^\infty\frac{d\xi}{\mathcal{H}(\xi)}
 \int_1^\xi d\eta\,
 \left[
 2\eta\partial_\eta V_2^{(0)}(\eta)
 +V_2^{(0)}(\eta)
 \right].
\label{app:V21-solution}
\end{equation}
Expanding the horizon-fixed integrand to linear order in $\alpha$ and performing the integrations gives
\begin{equation}
V_2^{(1)}
 =
 \frac{i\omega}{4r^2}
 \left[
 1+\log 2-6\alpha
 -2(1-2\alpha)\log r
 \right]
 +\mathcal{O}\left(\frac{\log r}{r^3}\right).
\label{app:V21-asymptotic}
\end{equation}

The equation determining $S_1^{(2)}$ is
\begin{equation}
(1+\alpha)
\left(
r^2\partial_r^2S_1^{(2)}
+3r\partial_rS_1^{(2)}
\right)
 =
 q^2\partial_rV_2^{(0)}.
\label{app:S12-equation-r}
\end{equation}
In terms of $z$,
\begin{equation}
\partial_z
\left(
z^3\partial_zS_1^{(2)}
\right)
 =
 \frac{q^2}{(1+\alpha)N}\,
 z\partial_zV_2^{(0)}.
\label{app:S12-equation-z}
\end{equation}
Its general boundary-vanishing solution can be written as
\begin{equation}
S_1^{(2)}(z)
 =
 -\int_z^\infty\frac{d\xi}{\xi^3}
 \left[
 C_1
 +\frac{q^2}{(1+\alpha)N}
 \int_1^\xi
 \eta\partial_\eta V_2^{(0)}(\eta)\,d\eta
 \right],
\label{app:S12-solution}
\end{equation}
where the normalizable homogeneous coefficient $C_1$ is fixed by the Landau-frame condition. The resulting near-boundary behaviour is
\begin{equation}
S_1^{(2)}
 =
 -\frac{(1-2\alpha)q^2}{4r^2}
 -\frac{(1-2\alpha)q^2\log r}{2r^2}
 +\mathcal{O}\left(\frac{\log r}{r^3}\right).
\label{app:S12-asymptotic}
\end{equation}

\paragraph{II: $\{S_2,V_1,V_3\}$.}

At zeroth order,
\begin{equation}
S_2^{(0)}=0,
\qquad
V_1^{(0)}=0.
\label{app:sectorII-zero}
\end{equation}
The equation for $V_3^{(0)}$ is
\begin{equation}
\partial_r
\left(
H\partial_rV_3^{(0)}
\right)
 =
 \frac{N}{r}.
\label{app:V30-equation-r}
\end{equation}
Horizon regularity fixes
\begin{equation}
H(r)\partial_rV_3^{(0)}(r)
 =
 N\log\left(\frac{r}{r_\text{h}}\right).
\label{app:V30-flux}
\end{equation}
Therefore,
\begin{equation}
V_3^{(0)}(r)
 =
 -N\int_r^\infty
 \frac{\log(x/r_\text{h})}{H(x)}\,dx,
\label{app:V30-r}
\end{equation}
or, equivalently,
\begin{equation}
V_3^{(0)}(z)
 =
 -R\int_z^\infty
 \frac{\log\xi}{\mathcal{H}(\xi)}\,d\xi.
\label{app:V30-z}
\end{equation}
Near the horizon,
\begin{equation}
\partial_zV_3^{(0)}(1)
 =
 \frac{R}{4(1-3\alpha)},
\end{equation}
so the solution is regular. At large $r$,
\begin{equation}
\begin{split}
V_3^{(0)}
 &=
 -\beta
 \left[
 \frac{\log(r/r_\text{h})}{2r^2}
 +\frac{1}{4r^2}
 \right]
 +\mathcal{O}\left(\frac{\log r}{r^3}\right)
 \\
 &=
 -\frac{(1-2\alpha)\log r}{2r^2}
 -\frac{1}{4r^2}
 +\mathcal{O}\left(\frac{\log r}{r^3}\right).
\end{split}
\label{app:V30-asymptotic}
\end{equation}
In the second line we used
\begin{equation}
\log r_\text{h}
 =
 \log(1-\alpha)
 =
 -\alpha+\mathcal{O}(\alpha^2).
\end{equation}

The first-order equation for $V_1$ is identical to the equation satisfied by $i\omega V_2^{(0)}$. With the same horizon and boundary conditions,
\begin{equation}
V_1^{(1)}=i\omega V_2^{(0)}.
\label{app:V11-relation}
\end{equation}
It follows that
\begin{equation}
V_1^{(1)}
 =
 -\frac{(1-\alpha)i\omega}{r}
 +\frac{(1-2\alpha)i\omega}{2r^2}
 +\mathcal{O}\left(\frac{1}{r^3}\right).
\label{app:V11-asymptotic}
\end{equation}

The equation for $S_2^{(1)}$ is
\begin{equation}
(1+\alpha)
\left(
r^2\partial_r^2S_2^{(1)}
+3r\partial_rS_2^{(1)}
\right)
 =
 -\partial_rV_1^{(1)}.
\label{app:S21-equation-r}
\end{equation}
Notice that the source involves $V_1^{(1)}$, rather than $V_1^{(0)}$, which vanishes. In terms of the horizon-fixed coordinate,
\begin{equation}
\partial_z
\left(
z^3\partial_zS_2^{(1)}
\right)
 =
 -\frac{z}{(1+\alpha)N}
 \partial_zV_1^{(1)}.
\label{app:S21-equation-z}
\end{equation}
Consequently,
\begin{equation}
S_2^{(1)}(z)
 =
 -\int_z^\infty\frac{d\xi}{\xi^3}
 \left[
 C_2
 -\frac{1}{(1+\alpha)N}
 \int_1^\xi
 \eta\partial_\eta V_1^{(1)}(\eta)\,d\eta
 \right],
\label{app:S21-solution}
\end{equation}
where $C_2$ is fixed by the Landau-frame condition. The resulting boundary expansion is
\begin{equation}
S_2^{(1)}
 =
 \frac{(1-2\alpha)i\omega}{4r^2}
 +\frac{(1-2\alpha)i\omega\log r}{2r^2}
 +\mathcal{O}\left(\frac{\log r}{r^3}\right).
\label{app:S21-asymptotic}
\end{equation}

At second hydrodynamic order, the equation for $V_1$ is
\begin{equation}
\partial_r
\left(
H\partial_rV_1^{(2)}
\right)
 =
 2i\omega r\partial_rV_1^{(1)}
 +i\omega V_1^{(1)}
 +\frac{Nq^2}{r}.
\label{app:V12-equation-r}
\end{equation}
The factor $R=1-\alpha$ multiplying the momentum source is essential. In the $z$ coordinate this becomes
\begin{equation}
\partial_z
\left(
\mathcal{H}\partial_zV_1^{(2)}
\right)
 =
 N\left[
 2i\omega z\partial_zV_1^{(1)}
 +i\omega V_1^{(1)}
 +\frac{q^2}{z}
 \right].
\label{app:V12-equation-z}
\end{equation}
The manifestly regular solution is
\begin{equation}
\begin{split}
V_1^{(2)}(z)
 =
 -R\int_z^\infty\frac{d\xi}{\mathcal{H}(\xi)}
 \int_1^\xi d\eta\,
 \left[
 2i\omega\eta\partial_\eta V_1^{(1)}(\eta)
 +i\omega V_1^{(1)}(\eta)
 +\frac{q^2}{\eta}
 \right].
\end{split}
\label{app:V12-solution}
\end{equation}
Its boundary expansion is
\begin{equation}
\begin{split}
V_1^{(2)}
 ={}&
 -\frac{1}{4r^2}
 \left[
 q^2+\omega^2(1+\log 2-6\alpha)
 \right]
 +\frac{1-2\alpha}{2r^2}
 (\omega^2-q^2)\log r
 +\mathcal{O}\left(\frac{\log r}{r^3}\right).
\end{split}
\label{app:V12-asymptotic}
\end{equation}

The same result follows more directly from the exact gauge-invariance identity
\begin{equation}
V_1=i\omega V_2+q^2V_3.
\label{app:gauge-identity-radial}
\end{equation}
Indeed, substituting \eqref{app:V20-asymptotic}, \eqref{app:V21-asymptotic}, and \eqref{app:V30-asymptotic} into \eqref{app:gauge-identity-radial} reproduces \eqref{app:V11-asymptotic} and \eqref{app:V12-asymptotic}.

\paragraph{III: $\{S_3,V_4\}$.}

At the first two orders,
\begin{equation}
S_3^{(0)}=S_3^{(1)}=0.
\label{app:S3-low-orders}
\end{equation}
The equation for $V_4^{(0)}$ is
\begin{equation}
\partial_r
\left(
H\partial_rV_4^{(0)}
\right)
 =
 -\frac{1}{2r^2}.
\label{app:V40-equation-r}
\end{equation}
Regularity at the physical horizon gives
\begin{equation}
H(r)\partial_rV_4^{(0)}(r)
 =
 \frac{1}{2}
 \left(
 \frac{1}{r}-\frac{1}{r_\text{h}}
 \right).
\label{app:V40-flux}
\end{equation}
Therefore,
\begin{equation}
V_4^{(0)}(r)
 =
 -\frac{1}{2}
 \int_r^\infty
 \frac{x^{-1}-r_\text{h}^{-1}}{H(x)}\,dx.
\label{app:V40-r}
\end{equation}
Equivalently,
\begin{equation}
V_4^{(0)}(z)
 =
 -\frac{1}{2R}
 \int_z^\infty
 \frac{\xi^{-1}-1}{\mathcal{H}(\xi)}\,d\xi.
\label{app:V40-z}
\end{equation}
This representation is real and regular throughout the entire radial interval $1\leq z<\infty$. In particular,
\begin{equation}
\partial_zV_4^{(0)}(1)
 =
 -\frac{1}{8N(1-3\alpha)}.
\end{equation}
The near-boundary expansion is
\begin{equation}
V_4^{(0)}
 =
 \frac{1}{4r^2}
 +\mathcal{O}\left(\frac{1}{r^3}\right).
\label{app:V40-asymptotic}
\end{equation}
The horizon-fixed representation \eqref{app:V40-z} avoids the branch-dependent quantities $\log(\alpha-1)$ and $\coth^{-1}(r^2)$ that arise if the fixed-$r$ kernel \eqref{app:nonuniform-kernel} is integrated across the shifted horizon.

At first order in frequency,
\begin{equation}
\partial_z
\left(
\mathcal{H}\partial_zV_4^{(1)}
\right)
 =
 Ni\omega
 \left(
 2z\partial_zV_4^{(0)}+V_4^{(0)}
 \right).
\label{app:V41-equation}
\end{equation}
The regular solution is
\begin{equation}
V_4^{(1)}(z)
 =
 -Ni\omega
 \int_z^\infty\frac{d\xi}{\mathcal{H}(\xi)}
 \int_1^\xi d\eta\,
 \left[
 2\eta\partial_\eta V_4^{(0)}(\eta)
 +V_4^{(0)}(\eta)
 \right].
\label{app:V41-solution}
\end{equation}
Evaluation of the horizon-fixed integrals gives
\begin{equation}
V_4^{(1)}
 =
 \frac{i\omega}{48r^2}
 \left[
 3\pi(1+2\alpha)-16\alpha
 \right]
 +\mathcal{O}\left(\frac{1}{r^3}\right).
\label{app:V41-asymptotic}
\end{equation}

The equation for $S_3^{(2)}$ is
\begin{equation}
(1+\alpha)
\left(
r^2\partial_r^2S_3^{(2)}
+3r\partial_rS_3^{(2)}
\right)
 =
 q^2\partial_rV_4^{(0)}.
\label{app:S32-equation-r}
\end{equation}
Equivalently,
\begin{equation}
\partial_z
\left(
z^3\partial_zS_3^{(2)}
\right)
 =
 \frac{q^2}{(1+\alpha)N}
 z\partial_zV_4^{(0)}.
\label{app:S32-equation-z}
\end{equation}
Thus,
\begin{equation}
S_3^{(2)}(z)
 =
 -\int_z^\infty\frac{d\xi}{\xi^3}
 \left[
 C_3
 +\frac{q^2}{(1+\alpha)R}
 \int_1^\xi
 \eta\partial_\eta V_4^{(0)}(\eta)\,d\eta
 \right].
\label{app:S32-solution}
\end{equation}
The Landau-frame condition fixes $C_3$ so that the normalizable $r^{-2}$ mode vanishes. Consequently,
\begin{equation}
S_3^{(2)}
 =
 \mathcal{O}\left(\frac{1}{r^3}\right).
\label{app:S32-asymptotic}
\end{equation}

\section*{Acknowledgements}
The authors thank Michael Lublinsky for his valuable comments and suggestions on the manuscript, and Yanyan Bu for assistance with implementing the spectral method used in the numerical calculations. A.N.K. acknowledges the support of Ben-Gurion University of the Negev, where most of this work was carried out during his postdoctoral appointment. He also acknowledges financial support from the Croatian Science Foundation through Project No. IP-2025-02-8625, \emph{Quantum aspects of gravity}.

\bibliography{BibTex}

@article{Bu:2015ame,
    author = "Bu, Yanyan and Lublinsky, Michael and Sharon, Amir",
    title = "{$U(1)$ current from the AdS/CFT: diffusion, conductivity and causality}",
    eprint = "1511.08789",
    archivePrefix = "arXiv",
    primaryClass = "hep-th",
    doi = "10.1007/JHEP04(2016)136",
    journal = "JHEP",
    volume = "04",
    pages = "136",
    year = "2016"
}

@article{Israel:1976tn,
    author = "Israel, W.",
    title = "{Nonstationary irreversible thermodynamics: A Causal relativistic theory}",
    doi = "10.1016/0003-4916(76)90064-6",
    journal = "Annals Phys.",
    volume = "100",
    pages = "310--331",
    year = "1976"
}

@article{Israel:1979wp,
    author = "Israel, W. and Stewart, J. M.",
    title = "{Transient relativistic thermodynamics and kinetic theory}",
    doi = "10.1016/0003-4916(79)90130-1",
    journal = "Annals Phys.",
    volume = "118",
    pages = "341--372",
    year = "1979"
}

@article{Maldacena:1997re,
    author = "Maldacena, Juan Martin",
    title = "{The Large N limit of superconformal field theories and supergravity}",
    eprint = "hep-th/9711200",
    archivePrefix = "arXiv",
    reportNumber = "HUTP-97-A097, HUTP-98-A097",
    doi = "10.4310/ATMP.1998.v2.n2.a1",
    journal = "Adv. Theor. Math. Phys.",
    volume = "2",
    pages = "231--252",
    year = "1998"
}

@article{Gubser:1998bc,
    author = "Gubser, S. S. and Klebanov, Igor R. and Polyakov, Alexander M.",
    title = "{Gauge theory correlators from noncritical string theory}",
    eprint = "hep-th/9802109",
    archivePrefix = "arXiv",
    reportNumber = "PUPT-1767",
    doi = "10.1016/S0370-2693(98)00377-3",
    journal = "Phys. Lett. B",
    volume = "428",
    pages = "105--114",
    year = "1998"
}

@article{Witten:1998qj,
    author = "Witten, Edward",
    title = "{Anti-de Sitter space and holography}",
    eprint = "hep-th/9802150",
    archivePrefix = "arXiv",
    reportNumber = "IASSNS-HEP-98-15",
    doi = "10.4310/ATMP.1998.v2.n2.a2",
    journal = "Adv. Theor. Math. Phys.",
    volume = "2",
    pages = "253--291",
    year = "1998"
}

@article{Policastro:2001yc,
    author = "Policastro, G. and Son, Dan T. and Starinets, Andrei O.",
    title = "{The Shear viscosity of strongly coupled N=4 supersymmetric Yang-Mills plasma}",
    eprint = "hep-th/0104066",
    archivePrefix = "arXiv",
    reportNumber = "NYU-TH-01-04-02, SNS-PH-01-05",
    doi = "10.1103/PhysRevLett.87.081601",
    journal = "Phys. Rev. Lett.",
    volume = "87",
    pages = "081601",
    year = "2001"
}

@article{Policastro:2002se,
    author = "Policastro, Giuseppe and Son, Dam T. and Starinets, Andrei O.",
    title = "{From AdS / CFT correspondence to hydrodynamics}",
    eprint = "hep-th/0205052",
    archivePrefix = "arXiv",
    reportNumber = "INT-PUB-02-32",
    doi = "10.1088/1126-6708/2002/09/043",
    journal = "JHEP",
    volume = "09",
    pages = "043",
    year = "2002"
}

@article{Kovtun:2004de,
    author = "Kovtun, P. and Son, Dan T. and Starinets, Andrei O.",
    title = "{Viscosity in strongly interacting quantum field theories from black hole physics}",
    eprint = "hep-th/0405231",
    archivePrefix = "arXiv",
    reportNumber = "INT-PUB-04-09, UW-PT-04-04",
    doi = "10.1103/PhysRevLett.94.111601",
    journal = "Phys. Rev. Lett.",
    volume = "94",
    pages = "111601",
    year = "2005"
}

@article{Iqbal:2008by,
    author = "Iqbal, Nabil and Liu, Hong",
    title = "{Universality of the hydrodynamic limit in AdS/CFT and the membrane paradigm}",
    eprint = "0809.3808",
    archivePrefix = "arXiv",
    primaryClass = "hep-th",
    reportNumber = "MIT-CTP-3983",
    doi = "10.1103/PhysRevD.79.025023",
    journal = "Phys. Rev. D",
    volume = "79",
    pages = "025023",
    year = "2009"
}

@article{Bhattacharyya:2007vjd,
    author = "Bhattacharyya, Sayantani and Hubeny, Veronika E and Minwalla, Shiraz and Rangamani, Mukund",
    title = "{Nonlinear Fluid Dynamics from Gravity}",
    eprint = "0712.2456",
    archivePrefix = "arXiv",
    primaryClass = "hep-th",
    reportNumber = "TIFR-TH-07-44, DCPT-07-73, NI07097",
    doi = "10.1088/1126-6708/2008/02/045",
    journal = "JHEP",
    volume = "02",
    pages = "045",
    year = "2008"
}

@article{Son:2007vk,
    author = "Son, Dam T. and Starinets, Andrei O.",
    title = "{Viscosity, Black Holes, and Quantum Field Theory}",
    eprint = "0704.0240",
    archivePrefix = "arXiv",
    primaryClass = "hep-th",
    reportNumber = "INT-PUB-07-02",
    doi = "10.1146/annurev.nucl.57.090506.123120",
    journal = "Ann. Rev. Nucl. Part. Sci.",
    volume = "57",
    pages = "95--118",
    year = "2007"
}

@article{Rangamani:2009xk,
    author = "Rangamani, Mukund",
    editor = "Uranga, A. M.",
    title = "{Gravity and Hydrodynamics: Lectures on the fluid-gravity correspondence}",
    eprint = "0905.4352",
    archivePrefix = "arXiv",
    primaryClass = "hep-th",
    reportNumber = "NSF-KITP-09-65",
    doi = "10.1088/0264-9381/26/22/224003",
    journal = "Class. Quant. Grav.",
    volume = "26",
    pages = "224003",
    year = "2009"
}

@article{Bu:2014sia,
    author = "Bu, Yanyan and Lublinsky, Michael",
    title = "{All order linearized hydrodynamics from fluid-gravity correspondence}",
    eprint = "1406.7222",
    archivePrefix = "arXiv",
    primaryClass = "hep-th",
    doi = "10.1103/PhysRevD.90.086003",
    journal = "Phys. Rev. D",
    volume = "90",
    number = "8",
    pages = "086003",
    year = "2014"
}

@article{Bu:2014ena,
    author = "Bu, Yanyan and Lublinsky, Michael",
    title = "{Linearized fluid/gravity correspondence: from shear viscosity to all order hydrodynamics}",
    eprint = "1409.3095",
    archivePrefix = "arXiv",
    primaryClass = "hep-th",
    doi = "10.1007/JHEP11(2014)064",
    journal = "JHEP",
    volume = "11",
    pages = "064",
    year = "2014"
}

@article{Lublinsky:2009kv,
    author = "Lublinsky, Michael and Shuryak, Edward",
    title = "{Improved Hydrodynamics from the AdS/CFT}",
    eprint = "0905.4069",
    archivePrefix = "arXiv",
    primaryClass = "hep-ph",
    doi = "10.1103/PhysRevD.80.065026",
    journal = "Phys. Rev. D",
    volume = "80",
    pages = "065026",
    year = "2009"
}

@article{Bu:2015bwa,
    author = "Bu, Yanyan and Lublinsky, Michael and Sharon, Amir",
    title = "{Hydrodynamics dual to Einstein-Gauss-Bonnet gravity: all-order gradient resummation}",
    eprint = "1504.01370",
    archivePrefix = "arXiv",
    primaryClass = "hep-th",
    doi = "10.1007/JHEP06(2015)162",
    journal = "JHEP",
    volume = "06",
    pages = "162",
    year = "2015"
}

@article{Bu:2015ika,
    author = "Bu, Yanyan and Lublinsky, Michael",
    title = "{Linearly resummed hydrodynamics in a weakly curved spacetime}",
    eprint = "1502.08044",
    archivePrefix = "arXiv",
    primaryClass = "hep-th",
    doi = "10.1007/JHEP04(2015)136",
    journal = "JHEP",
    volume = "04",
    pages = "136",
    year = "2015"
}

@article{Crossley:2015evo,
    author = "Crossley, Michael and Glorioso, Paolo and Liu, Hong",
    title = "{Effective field theory of dissipative fluids}",
    eprint = "1511.03646",
    archivePrefix = "arXiv",
    primaryClass = "hep-th",
    reportNumber = "MIT-CTP-4734",
    doi = "10.1007/JHEP09(2017)095",
    journal = "JHEP",
    volume = "09",
    pages = "095",
    year = "2017"
}

@article{Karch:2002sh,
    author = "Karch, Andreas and Katz, Emanuel",
    title = "{Adding flavor to AdS / CFT}",
    eprint = "hep-th/0205236",
    archivePrefix = "arXiv",
    reportNumber = "UW-PT-02-10",
    doi = "10.1088/1126-6708/2002/06/043",
    journal = "JHEP",
    volume = "06",
    pages = "043",
    year = "2002"
}

@article{Kovtun:2003wp,
    author = "Kovtun, Pavel and Son, Dam T. and Starinets, Andrei O.",
    title = "{Holography and hydrodynamics: Diffusion on stretched horizons}",
    eprint = "hep-th/0309213",
    archivePrefix = "arXiv",
    reportNumber = "UW-PT-03-21, INT-PUB-03-17",
    doi = "10.1088/1126-6708/2003/10/064",
    journal = "JHEP",
    volume = "10",
    pages = "064",
    year = "2003"
}

@article{Horowitz:2008bn,
    author = "Horowitz, Gary T. and Roberts, Matthew M.",
    title = "{Holographic Superconductors with Various Condensates}",
    eprint = "0810.1077",
    archivePrefix = "arXiv",
    primaryClass = "hep-th",
    doi = "10.1103/PhysRevD.78.126008",
    journal = "Phys. Rev. D",
    volume = "78",
    pages = "126008",
    year = "2008"
}

@article{Myers:2007we,
    author = "Myers, Robert C. and Starinets, Andrei O. and Thomson, Rowan M.",
    title = "{Holographic spectral functions and diffusion constants for fundamental matter}",
    eprint = "0706.0162",
    archivePrefix = "arXiv",
    primaryClass = "hep-th",
    doi = "10.1088/1126-6708/2007/11/091",
    journal = "JHEP",
    volume = "11",
    pages = "091",
    year = "2007"
}

@article{Taylor:2000xw,
    author = "Taylor, Marika",
    title = "{More on counterterms in the gravitational action and anomalies}",
    eprint = "hep-th/0002125",
    archivePrefix = "arXiv",
    month = "2",
    year = "2000"
}

@article{Bianchi:2001kw,
    author = "Bianchi, Massimo and Freedman, Daniel Z. and Skenderis, Kostas",
    title = "{Holographic renormalization}",
    eprint = "hep-th/0112119",
    archivePrefix = "arXiv",
    reportNumber = "MIT-CTP-3166, PUTP-1999, DAMTP-2001-63, ROM2F-2001-30",
    doi = "10.1016/S0550-3213(02)00179-7",
    journal = "Nucl. Phys. B",
    volume = "631",
    pages = "159--194",
    year = "2002"
}

@article{Sahoo:2010sp,
    author = "Sahoo, Bindusar and Yee, Ho-Ung",
    title = "{Electrified plasma in AdS/CFT correspondence}",
    eprint = "1004.3541",
    archivePrefix = "arXiv",
    primaryClass = "hep-th",
    doi = "10.1007/JHEP11(2010)095",
    journal = "JHEP",
    volume = "11",
    pages = "095",
    year = "2010"
}

@article{Son:2002sd,
    author = "Son, Dam T. and Starinets, Andrei O.",
    title = "{Minkowski space correlators in AdS / CFT correspondence: Recipe and applications}",
    eprint = "hep-th/0205051",
    archivePrefix = "arXiv",
    reportNumber = "INT-PUB-02-34",
    doi = "10.1088/1126-6708/2002/09/042",
    journal = "JHEP",
    volume = "09",
    pages = "042",
    year = "2002"
}

@article{Brihaye:2008kh,
    author = "Brihaye, Yves and Radu, Eugen",
    title = "{Five-dimensional rotating black holes in Einstein-Gauss-Bonnet theory}",
    eprint = "0801.1021",
    archivePrefix = "arXiv",
    primaryClass = "hep-th",
    doi = "10.1016/j.physletb.2008.02.005",
    journal = "Phys. Lett. B",
    volume = "661",
    pages = "167--174",
    year = "2008"
}

@article{Myers:1987yn,
    author = "Myers, Robert C.",
    title = "{Higher Derivative Gravity, Surface Terms and String Theory}",
    reportNumber = "NSF-ITP-87-16",
    doi = "10.1103/PhysRevD.36.392",
    journal = "Phys. Rev. D",
    volume = "36",
    pages = "392",
    year = "1987"
}

@article{Davis:2002gn,
    author = "Davis, Stephen C.",
    title = "{Generalized Israel junction conditions for a Gauss-Bonnet brane world}",
    eprint = "hep-th/0208205",
    archivePrefix = "arXiv",
    doi = "10.1103/PhysRevD.67.024030",
    journal = "Phys. Rev. D",
    volume = "67",
    pages = "024030",
    year = "2003"
}

@article{Brigante:2007nu,
    author = "Brigante, Mauro and Liu, Hong and Myers, Robert C. and Shenker, Stephen and Yaida, Sho",
    title = "{Viscosity Bound Violation in Higher Derivative Gravity}",
    eprint = "0712.0805",
    archivePrefix = "arXiv",
    primaryClass = "hep-th",
    reportNumber = "CAS-KITPC-ITP-025, MIT-CTP-3918, SU-ITP-07-22",
    doi = "10.1103/PhysRevD.77.126006",
    journal = "Phys. Rev. D",
    volume = "77",
    pages = "126006",
    year = "2008"
}

@article{Brigante:2008gz,
    author = "Brigante, Mauro and Liu, Hong and Myers, Robert C. and Shenker, Stephen and Yaida, Sho",
    title = "{The Viscosity Bound and Causality Violation}",
    eprint = "0802.3318",
    archivePrefix = "arXiv",
    primaryClass = "hep-th",
    reportNumber = "SU-ITP-08-02, MIT-CTP-3929",
    doi = "10.1103/PhysRevLett.100.191601",
    journal = "Phys. Rev. Lett.",
    volume = "100",
    pages = "191601",
    year = "2008"
}

@article{Buchel:2009tt,
    author = "Buchel, Alex and Myers, Robert C.",
    title = "{Causality of Holographic Hydrodynamics}",
    eprint = "0906.2922",
    archivePrefix = "arXiv",
    primaryClass = "hep-th",
    reportNumber = "UWO-TH-09-11",
    doi = "10.1088/1126-6708/2009/08/016",
    journal = "JHEP",
    volume = "08",
    pages = "016",
    year = "2009"
}

@article{Buchel:2009sk,
    author = "Buchel, Alex and Escobedo, Jorge and Myers, Robert C. and Paulos, Miguel F. and Sinha, Aninda and Smolkin, Michael",
    title = "{Holographic GB gravity in arbitrary dimensions}",
    eprint = "0911.4257",
    archivePrefix = "arXiv",
    primaryClass = "hep-th",
    reportNumber = "UWO-TH-09-16",
    doi = "10.1007/JHEP03(2010)111",
    journal = "JHEP",
    volume = "03",
    pages = "111",
    year = "2010"
}

@article{Bu:2016oba,
    author = "Bu, Yanyan and Lublinsky, Michael and Sharon, Amir",
    title = "{Anomalous transport from holography: Part I}",
    eprint = "1608.08595",
    archivePrefix = "arXiv",
    primaryClass = "hep-th",
    doi = "10.1007/JHEP11(2016)093",
    journal = "JHEP",
    volume = "11",
    pages = "093",
    year = "2016"
}

@article{Kubo:1957mj,
    author = "Kubo, Ryogo",
    title = "{Statistical mechanical theory of irreversible processes. 1. General theory and simple applications in magnetic and conduction problems}",
    doi = "10.1143/JPSJ.12.570",
    journal = "J. Phys. Soc. Jap.",
    volume = "12",
    pages = "570--586",
    year = "1957"
}

@book{forster2018hydrodynamic,
  title={Hydrodynamic Fluctuations, Broken Symmetry, And Correlation Functions},
  author={Forster, D.},
  isbn={9780429973314},
  url={https://books.google.co.in/books?id=IwpQDwAAQBAJ},
  year={2018},
  publisher={CRC Press}
}

@article{Kovtun:2012rj,
    author = "Kovtun, Pavel",
    title = "{Lectures on hydrodynamic fluctuations in relativistic theories}",
    eprint = "1205.5040",
    archivePrefix = "arXiv",
    primaryClass = "hep-th",
    doi = "10.1088/1751-8113/45/47/473001",
    journal = "J. Phys. A",
    volume = "45",
    pages = "473001",
    year = "2012"
}

@book{Romatschke:2017ejr,
    author = "Romatschke, Paul and Romatschke, Ulrike",
    title = "{Relativistic Fluid Dynamics In and Out of Equilibrium}",
    eprint = "1712.05815",
    archivePrefix = "arXiv",
    primaryClass = "nucl-th",
    doi = "10.1017/9781108651998",
    isbn = "978-1-108-48368-1, 978-1-108-75002-8",
    publisher = "Cambridge University Press",
    series = "Cambridge Monographs on Mathematical Physics",
    month = "5",
    year = "2019"
}

@article{Janik:2005zt,
    author = "Janik, Romuald A. and Peschanski, Robert B.",
    title = "{Asymptotic perfect fluid dynamics as a consequence of Ads/CFT}",
    eprint = "hep-th/0512162",
    archivePrefix = "arXiv",
    doi = "10.1103/PhysRevD.73.045013",
    journal = "Phys. Rev. D",
    volume = "73",
    pages = "045013",
    year = "2006"
}

@article{Banerjee:2008th,
    author = "Banerjee, Nabamita and Bhattacharya, Jyotirmoy and Bhattacharyya, Sayantani and Dutta, Suvankar and Loganayagam, R. and Surowka, P.",
    title = "{Hydrodynamics from charged black branes}",
    eprint = "0809.2596",
    archivePrefix = "arXiv",
    primaryClass = "hep-th",
    doi = "10.1007/JHEP01(2011)094",
    journal = "JHEP",
    volume = "01",
    pages = "094",
    year = "2011"
}

@article{Erdmenger:2008rm,
    author = "Erdmenger, Johanna and Haack, Michael and Kaminski, Matthias and Yarom, Amos",
    title = "{Fluid dynamics of R-charged black holes}",
    eprint = "0809.2488",
    archivePrefix = "arXiv",
    primaryClass = "hep-th",
    reportNumber = "LMU-ASC-48-08, MPP-2008-116",
    doi = "10.1088/1126-6708/2009/01/055",
    journal = "JHEP",
    volume = "01",
    pages = "055",
    year = "2009"
}

@article{Haack:2008xx,
    author = "Haack, Michael and Yarom, Amos",
    title = "{Universality of second order transport coefficients from the gauge-string duality}",
    eprint = "0811.1794",
    archivePrefix = "arXiv",
    primaryClass = "hep-th",
    reportNumber = "PUPT-2286, LMU-ASC-53-08",
    doi = "10.1016/j.nuclphysb.2008.12.028",
    journal = "Nucl. Phys. B",
    volume = "813",
    pages = "140--155",
    year = "2009"
}

@article{Stewart1977,
  author  = {Stewart, J. M.},
  title   = {On Transient Relativistic Thermodynamics and Kinetic Theory},
  journal = {Proceedings of the Royal Society of London. Series A, Mathematical and Physical Sciences},
  volume  = {357},
  number  = {1688},
  pages   = {59--75},
  year    = {1977}
}

@article{Baier:2007ix,
    author = "Baier, Rudolf and Romatschke, Paul and Son, Dam Thanh and Starinets, Andrei O. and Stephanov, Mikhail A.",
    title = "{Relativistic viscous hydrodynamics, conformal invariance, and holography}",
    eprint = "0712.2451",
    archivePrefix = "arXiv",
    primaryClass = "hep-th",
    reportNumber = "BI-TP-2007-29, INT-PUB-07-45, SHEP-07-47",
    doi = "10.1088/1126-6708/2008/04/100",
    journal = "JHEP",
    volume = "04",
    pages = "100",
    year = "2008"
}

@article{Kovtun:2019hdm,
    author = "Kovtun, Pavel",
    title = "{First-order relativistic hydrodynamics is stable}",
    eprint = "1907.08191",
    archivePrefix = "arXiv",
    primaryClass = "hep-th",
    doi = "10.1007/JHEP10(2019)034",
    journal = "JHEP",
    volume = "10",
    pages = "034",
    year = "2019"
}

@article{Mori:1965oqj,
    author = "Mori, Hazime",
    title = "{Transport, Collective Motion, and Brownian Motion}",
    doi = "10.1143/PTP.33.423",
    journal = "Prog. Theor. Phys.",
    volume = "33",
    number = "3",
    pages = "423--455",
    year = "1965"
}

@article{Heller:2015dha,
    author = "Heller, Michal P. and Spalinski, Michal",
    title = "{Hydrodynamics Beyond the Gradient Expansion: Resurgence and Resummation}",
    eprint = "1503.07514",
    archivePrefix = "arXiv",
    primaryClass = "hep-th",
    doi = "10.1103/PhysRevLett.115.072501",
    journal = "Phys. Rev. Lett.",
    volume = "115",
    number = "7",
    pages = "072501",
    year = "2015"
}

@article{Romatschke:2017vte,
    author = "Romatschke, Paul",
    title = "{Relativistic Fluid Dynamics Far From Local Equilibrium}",
    eprint = "1704.08699",
    archivePrefix = "arXiv",
    primaryClass = "hep-th",
    doi = "10.1103/PhysRevLett.120.012301",
    journal = "Phys. Rev. Lett.",
    volume = "120",
    number = "1",
    pages = "012301",
    year = "2018"
}

@article{Bu:2016vum,
    author = "Bu, Yanyan and Lublinsky, Michael and Sharon, Amir",
    title = "{Anomalous transport from holography: Part II}",
    eprint = "1609.09054",
    archivePrefix = "arXiv",
    primaryClass = "hep-th",
    doi = "10.1140/epjc/s10052-017-4762-4",
    journal = "Eur. Phys. J. C",
    volume = "77",
    number = "3",
    pages = "194",
    year = "2017"
}

@article{Bu:2018psl,
    author = "Bu, Yanyan and Demircik, Tuna and Lublinsky, Michael",
    title = "{Nonlinear chiral transport from holography}",
    eprint = "1807.08467",
    archivePrefix = "arXiv",
    primaryClass = "hep-th",
    doi = "10.1007/JHEP01(2019)078",
    journal = "JHEP",
    volume = "01",
    pages = "078",
    year = "2019"
}

@article{Bu:2018drd,
    author = "Bu, Yanyan and Demircik, Tuna and Lublinsky, Michael",
    title = "{Gradient resummation for nonlinear chiral transport: an insight from holography}",
    eprint = "1807.11908",
    archivePrefix = "arXiv",
    primaryClass = "hep-th",
    doi = "10.1140/epjc/s10052-019-6576-z",
    journal = "Eur. Phys. J. C",
    volume = "79",
    number = "1",
    pages = "54",
    year = "2019"
}

@article{Bu:2019mow,
    author = "Bu, Yanyan and Demircik, Tuna and Lublinsky, Michael",
    title = "{Chiral transport in strong fields from holography}",
    eprint = "1903.00896",
    archivePrefix = "arXiv",
    primaryClass = "hep-th",
    doi = "10.1007/JHEP05(2019)071",
    journal = "JHEP",
    volume = "05",
    pages = "071",
    year = "2019"
}

@article{Bu:2020jfo,
    author = "Bu, Yanyan and Demircik, Tuna and Lublinsky, Michael",
    title = "{All order effective action for charge diffusion from Schwinger-Keldysh holography}",
    eprint = "2012.08362",
    archivePrefix = "arXiv",
    primaryClass = "hep-th",
    doi = "10.1007/JHEP05(2021)187",
    journal = "JHEP",
    volume = "05",
    pages = "187",
    year = "2021"
}

@article{Liu:2018kfw,
    author = "Liu, Hong and Glorioso, Paolo",
    title = "{Lectures on non-equilibrium effective field theories and fluctuating hydrodynamics}",
    eprint = "1805.09331",
    archivePrefix = "arXiv",
    primaryClass = "hep-th",
    reportNumber = "MIT-CTP/5018; EFI-18-8, MIT-CTP-5018, EFI-18-8",
    doi = "10.22323/1.305.0008",
    journal = "PoS",
    volume = "TASI2017",
    pages = "008",
    year = "2018"
}

@article{Haehl:2014zda,
    author = "Haehl, Felix M. and Loganayagam, R. and Rangamani, Mukund",
    title = "{The eightfold way to dissipation}",
    eprint = "1412.1090",
    archivePrefix = "arXiv",
    primaryClass = "hep-th",
    reportNumber = "DCPT-14-65",
    doi = "10.1103/PhysRevLett.114.201601",
    journal = "Phys. Rev. Lett.",
    volume = "114",
    pages = "201601",
    year = "2015"
}

@article{Kats:2007mq,
    author = "Kats, Yevgeny and Petrov, Pavel",
    title = "{Effect of curvature squared corrections in AdS on the viscosity of the dual gauge theory}",
    eprint = "0712.0743",
    archivePrefix = "arXiv",
    primaryClass = "hep-th",
    doi = "10.1088/1126-6708/2009/01/044",
    journal = "JHEP",
    volume = "01",
    pages = "044",
    year = "2009"
}

@article{Hu:2011ze,
    author = "Hu, Ya-Peng and Sun, Peng and Zhang, Jian-Hui",
    title = "{Hydrodynamics with conserved current via AdS/CFT correspondence in the Maxwell-Gauss-Bonnet gravity}",
    eprint = "1103.3773",
    archivePrefix = "arXiv",
    primaryClass = "hep-th",
    doi = "10.1103/PhysRevD.83.126003",
    journal = "Phys. Rev. D",
    volume = "83",
    pages = "126003",
    year = "2011"
}

@article{Hu:2011qa,
    author = "Hu, Ya-Peng and Park, Chanyong",
    title = "{Chern-Simons effect on the dual hydrodynamics in the Maxwell-Gauss-Bonnet gravity}",
    eprint = "1112.4227",
    archivePrefix = "arXiv",
    primaryClass = "hep-th",
    doi = "10.1016/j.physletb.2012.07.008",
    journal = "Phys. Lett. B",
    volume = "714",
    pages = "324--330",
    year = "2012"
}

@article{Grozdanov:2017kyl,
    author = "Grozdanov, Sa{\v{s}}o and Poovuttikul, Napat",
    title = "{Generalised global symmetries in holography: magnetohydrodynamic waves in a strongly interacting plasma}",
    eprint = "1707.04182",
    archivePrefix = "arXiv",
    primaryClass = "hep-th",
    doi = "10.1007/JHEP04(2019)141",
    journal = "JHEP",
    volume = "04",
    pages = "141",
    year = "2019"
}

@article{Mas:2008qs,
    author = "Mas, Javier and Shock, Jonathan P and Tarrio, Javier",
    title = "{A Note on conductivity and charge diffusion in holographic flavour systems}",
    eprint = "0811.1750",
    archivePrefix = "arXiv",
    primaryClass = "hep-th",
    doi = "10.1088/1126-6708/2009/01/025",
    journal = "JHEP",
    volume = "01",
    pages = "025",
    year = "2009"
}

@article{Baggioli:2024zfq,
    author = "Baggioli, Matteo and Bu, Yanyan and Sun, Xiyang",
    title = "{Chiral anomalous magnetohydrodynamics in action: effective field theory and holography}",
    eprint = "2412.02361",
    archivePrefix = "arXiv",
    primaryClass = "hep-th",
    doi = "10.1007/JHEP04(2025)126",
    journal = "JHEP",
    volume = "04",
    pages = "126",
    year = "2025"
}

@article{Lublinsky:2026wvt,
    author = "Lublinsky, Michael and Tzarfati, Hadas",
    title = "{Chiral Plasma under Strong Magnetic Fields: A Holographic Analysis of Transport Phenomena}",
    eprint = "2606.10689",
    archivePrefix = "arXiv",
    primaryClass = "hep-th",
    month = "6",
    year = "2026"
}

@article{Lublinsky:2007mm,
  author = {Lublinsky, Michael and Shuryak, Edward},
  title = {{How much entropy is produced in strongly coupled Quark-Gluon Plasma (sQGP) by dissipative effects?}},
  eprint = {0704.1647},
  archiveprefix = {arXiv},
  primaryclass = {hep-ph},
  doi = {10.1103/PhysRevC.76.021901},
  journal = {Phys. Rev. C},
  volume = {76},
  pages = {021901},
  year = {2007},
}

@article{Lublinsky:2011cw,
  author = {Lublinsky, M. and Shuryak, E.},
  title = {{Universal hydrodynamics and charged hadron multiplicity at the LHC}},
  eprint = {1108.3972},
  archiveprefix = {arXiv},
  primaryclass = {hep-ph},
  doi = {10.1103/PhysRevC.84.061901},
  journal = {Phys. Rev. C},
  volume = {84},
  pages = {061901},
  year = {2011},
}
\end{document}